\documentclass[fleqn,usenatbib]{mnras}

\usepackage{newtxtext,newtxmath}

\usepackage[T1]{fontenc}
\usepackage{ae,aecompl}

\usepackage{graphicx}	
\usepackage{amsmath}	
\usepackage{amssymb}	

\title[A non evolving MS in slope and scatter]{The main sequence of star forming galaxies II. A non evolving slope at the high mass end}

\author[P.Popesso et al.]{
P. Popesso,$^{1}$\thanks{E-mail: paola.popesso@tum.de}
L. Morselli,$^{2}$
A. Concas,$^{3}$
C. Schreiber,$^{4}$
G. Rodighiero,$^{2}$
\newauthor
G. Cresci,$^{5}$
S. Belli,$^{6}$
O. Ilbert,$^{7}$
G. Erfanianfar,$^{6}$
C. Mancini,$^{2}$
H. Inami,$^{8}$
M. Dickinson,$^{9}$
\newauthor
M. Pannella,$^{10}$
D. Elbaz$^{11}$
\\
$^{1}$Excellence Cluster Universe, Boltzmannstrasse 2, 85748, Garching bei M\"unchen, Germany\\
$^{2}$Dipartimento di Fisica e Astronomia, Universit\'a di Padova, Vicolo dell' Osservatorio 3, 35122, Padova, Italy\\
$^{3}$Kavli Institute for Cosmology, University of Cambridge, Madingley Road, Cambridge CB3 0HA \\
$^{4}$Leiden Observatory, Leiden University, 2300 RA, Leiden, The Netherlands\\
$^{5}$INAF-Osservatorio Astronomico di Arcetri, Largo Enrico Fermi 5, 50125, Firenze, Italy\\
$^{6}$Max Planck f\"ur Extraterrestrische Physik, Giessenbachstrasse 1, 85478, Garching bei M\"unchen, Germany\\
$^{7}$Laboratoire d'Astrophysique de Marseille, 38 rue Frederic Joliot Curie, 13388 Marseille, France \\
$^{8}$Universit\'e Lyon 1, 9 Avenue Charles Andr\'e, 69561, Saint Genis Laval, France\\
$^{9}$National Optical Astronomy Observatories, 950 N Cherry Avenue, Tucson, Arizona 85719, USA \\
$^{10}$Faculty of Physics, Ludwig-Maximilians-Universit\"at, Scheinerstr. 1, D-81679 Munich, Germany\\
$^{11}$Laboratoire AIM-Paris-Saclay, CEA/DRF/Irfu - CNRS - Universit'e Paris Diderot, CEA-Saclay, 91191 Gif-sur-Yvette, France
}

\date{Accepted XXX. Received YYY; in original form ZZZ}

\pubyear{2019}

\begin{document}
\label{firstpage}
\pagerange{\pageref{firstpage}--\pageref{lastpage}}
\maketitle

\begin{abstract}
By using the deepest available mid and far infrared surveys in the CANDELS, GOODS and COSMOS fields we study the evolution of the Main Sequence (MS) of star forming galaxies (SFGs) from z $\sim$ 0 to $\sim$ 2.5 at stellar masses larger than $10^{10} M_{\odot}$. The MS slope and scatter are consistent with a rescaled version of the local relation and distribution, shifted at higher values of SFR according to $\propto(1+z)^{3.2}$. The relation exhibits a bending at the high mass end and a slightly increasing scatter as a function of the stellar mass. We show that the previously reported evolution of the MS slope, in the considered mass and redshift range, is due to a selection effect. The distribution of galaxies in the MS region at fixed stellar mass is well represented by a single log-normal distribution at all redshifts and masses, with starburst galaxies (SBs) occupying the tail at high SFR.
\end{abstract}




\section{Introduction}

How and when the stellar mass content of galaxies is assembled is still one of the main questions of galaxy evolution studies. The existence of a very tight relation between the galaxy star formation rate (SFR) and the stellar mass (M$_*$) suggests that most galaxies form their stars at a level mainly dictated by their stellar masses and regulated by secular processes. Such relation, called the Main Sequence (MS) of star forming galaxies (SFGs), is in place from redshift $\sim$ 0 up to $\sim$ 4 \citep{2004MNRAS.351.1151B, 2007ApJS..173..267S, 2007ApJ...660L..47N,2007A&A...468...33E,2007ApJ...670..156D,2009MNRAS.393..406C,2009ApJ...698L.116P,2009A&A...504..751S,2010MNRAS.405.2279O,2010ApJ...714.1740M,2011ApJ...739L..40R,2011A&A...533A.119E,2011ApJ...730...61K,2011ApJ...738...69S,2012ApJ...754L..29W,2012ApJ...750..120Z,2012ApJ...752...66L,2012ApJ...744..154R,Salmi:2012hm,2013ApJ...767...50M,2013ApJ...777L...8K,2014MNRAS.437.3516S,2014ApJ...791L..25S,2014ApJS..214...15S,2014ApJ...795..104W,2015ApJ...804..149S,2015A&A...575A..74S,2015A&A...581A..54T,2015ApJ...801...80L,2016ApJ...820L...1K,2016MNRAS.455.2839E,2017ApJ...847...76S,2018A&A...615A.146P}. It is considered one of the most useful tools in astrophysics to study the evolution of the star formation activity in galaxies \citep{2017ApJ...844...45O}. 

The evolution of normalization, slope and scatter of the relation have been largely studied in the past decade. Most of the studies agree in finding that slope and scatter evolve moderately, while the normalization declines significantly but smoothly as a function of cosmic time as $\propto (1+z)^{\gamma}$, with $\gamma$ in the range 1.9-3.5, likely depending on the stellar mass \citep{2014ApJS..214...15S,2015ApJ...811L..12W,2015A&A...575A..74S,2015A&A...579A...2I,2018A&A...615A.146P}. This would suggest that prior to the shutdown of star formation, the galaxy star formation history declines smoothly on mass-dependent timescales \citep[see][]{2004Natur.428..625H}, rather than being driven by stochastic events like major mergers and starbursts \citep[SBs,][]{2017ApJ...844...45O}. 

The exact shape of the relation and its evolution, instead, are still a matter of debate. Several studies point to a power law shape, SFR $\propto$ M$_*^{\alpha}$, with an intrinsic scatter of about 0.2-0.3 dex for moderate to relatively low stellar mass galaxies, both in the local Universe \citep{2010AAS...21522909P,2015ApJ...801L..29R} and at high redshift \citep{2014ApJS..214...15S,2014MNRAS.443...19R,2016ApJ...820L...1K,2018A&A...615A.146P}. Speagle et al. (2014) combine and cross-calibrate 25 MS determinations available in literature and find that $\alpha\sim 0.84$ with almost negligible evolution across cosmic time. Other works suggest that the relation exhibits a curvature towards the high mass end with a flatter slope with respect to the low mass regime at low \citep{2019MNRAS.483.3213P} and high redshift \citep{2014ApJ...795..104W,2015A&A...575A..74S,2015ApJ...801...80L,2015A&A...581A..54T,2016ApJ...817..118T}. While at low stellar masses the relation seems to be relatively stable as a function of redshift across all analyses, these studies suggest that the curvature and stellar mass of the turn-over evolve with time. This would lead to a more significant bending of the MS in the local Universe with respect to the much steeper MS observed at $z\sim 2$ \citep{2014ApJ...795..104W,2015A&A...575A..74S,2015ApJ...801...80L,2015A&A...581A..54T,2016ApJ...817..118T}.

The large spread of results at the high mass end can be explained by the distribution of galaxies in the log(SFR)-log(M$_*$) plane. The distinction between MS, quiescence region and the valley in-between, becomes progressively less clear towards high stellar masses, where the MS is not isolated and unmistakably identifiable \citep{2019MNRAS.483.3213P}. Additionally, low mass SF galaxies are predominately blue disks with a flat spatial specific SFR distribution \citep{2011ApJ...742...96W,2015ApJ...811L..12W, 2017A&A...597A..97M,2019A&A...626A..61M}, while high mass systems exhibit a mixture of morphologies with a suppressed SF activity towards the center and a larger spread of colors and integrated SFRs at low \citep{2017A&A...597A..97M,2018MNRAS.477.3014B} and high redshift \citep{2012ApJ...754L..29W,2015ApJ...811L..12W}. All together, these aspects make it difficult to unambiguously identify massive SFGs, and introduce several selection effects. 

One solution, largely exploited in literature, is to select SFGs at any mass according to their colors. This technique is applied with slightly different color cut definitions in the BzK \citep{2007ApJ...670..156D,2010A&A...518L..25R,2014MNRAS.443...19R}, rest-frame (U-V) $-$ (V-J) and (M$_{NUV}$-M$_R$) $-$ (M$_R$-M$_J$) absolute color diagrams. However, as shown in \cite{2015A&A...575A..74S} such selection is able to capture the whole FIR selected galaxy population, observed with {\it{Herschel}} PACS, below $\sim 10^{10.5}$ $M_{\odot}$, but it misses a fraction from 30 to 50\% of the most massive infrared selected galaxies, in particular at high redshift. Similarly, \cite{2015MNRAS.453.2540J} show that a strict 4000 \AA \hspace{0.05cm} break cut leads to a power law MS, while a less strict cut includes lower SFR galaxies at the high mass end and forms a relation with a turn-over. 

To solve this problem, \cite{2015ApJ...801L..29R} define a clean method to identify the exact location of the relation as the ridge line connecting the peaks of the 3D number density distribution of galaxies over the log(SFR)-log(M$_*$) plane. Such definition takes into account only the distribution of galaxies in the log(SFR)-log(M$_*$) plane without any SF galaxy pre-selection. Following this procedure, \cite{2019MNRAS.483.3213P} show that the distribution of local galaxies in the MS locus is log-normal and the peak can be unambiguously identified over the stellar mass range from $10^{10}$ to $10^{11.5}$ M$_{\odot}$. This leads to a local MS relation with a turn-over at $\sim10^{10.4-10.6}$ M$_{\odot}$. The effect of a color selection, e.g. as in Peng et al. (2010), is to get rid of half of the log-normal distribution at lower SFR and high mass, resulting in a biased power law MS. Similar analyses in literature show that the distribution of galaxies in the MS region is log-normal and with a clear peak over the same stellar mass range up to z$\sim$2 \citep{2011ApJ...739L..40R,2015A&A...575A..74S}. This would suggest that the MS at the high mass end can be unambiguously identified with this method up to high redshift, as long as the MS locus is sampled with high completeness in SFR and stellar mass. 

In this paper we take advantage of the deepest available {\it{Spitzer}} and {\it{Herschel}} mid- and far-infrared surveys ever conducted, to sample the MS region with high completeness and a robust SFR estimator over the stellar mass range $10^{10}$-$10^{11.5}$ $M_{\odot}$. We study the distribution of galaxies in the MS locus, to identify the relation up to $\sim 2.5$. Differently from all previous works, we tackle the issue by testing the null hypothesis that the MS does not evolve from redshift $\sim$ 0 to 2.5 in slope and scatter but only in its normalization, as suggested in literature for the low mass end of the relation \citep{2014ApJS..214...15S,2015A&A...575A..74S,2016ApJ...820L...1K,2017ApJ...847...76S}. Our approach has the advantage of reducing the analysis to a single parameter fit (normalisation), because we assume that the well constrained slope and scatter of the local MS are valid up to z$\sim$2.5. To this purpose we use as local benchmark the distribution of the WISE 22 $\mu$m selected galaxies in the local Universe studied by \cite{2019MNRAS.483.3213P} and test the validity of the null hypothesis by combining the depth of the GOODS and CANDELS {\it{Spitzer}} and {\it{Herschel}} surveys \citep[][Inami et al. in prep.]{2013A&A...553A.132M} with the 10 times larger sky coverage of the shallower {\it{Spitzer}} and {\it{Herschel}} observations of the COSMOS field as part of the PEP survey \citep{2011A&A...532A..90L}.

\begin{table*}
\caption{Catalog depths for each field. \label{TAB:depth}}
\begin{center}
\begin{tabular}{lcccc}
    \hline
    \hline \\[-2.5mm]
    Field & Area & $24\,\mu{\rm m}$ & $100\,\mu{\rm m}$ & $160\,\mu{\rm m}$  \\
          &                  & $\mu{\rm Jy}$ ($3\sigma$) & ${\rm mJy}$ ($3\sigma)$ & ${\rm mJy}$ ($3\sigma)$ \\
    \hline \\[-2.5mm]
       GOODS$+$CANDELS    & &    &  &  \\
    \hline \\[-2.5mm]
    GOODS-N     & $168\,{\rm arcmin}^2$ & $21$       & $1.1$  & $2.7$ \\
    GOODS-S     & $184\,{\rm arcmin}^2$ & $20$       & $0.8$  & $2.4$ \\
    UDS    & $202\,{\rm arcmin}^2$ & $40$       & $1.7$  & $3.9$  \\[1pt]
    COSMOS & $208\,{\rm arcmin}^2$ &  $27$--$40$ & $1.5$  & $3.1$ \\
       \hline 
          \hline \\[-2.5mm]
   COSMOS-PEP & $2\,\deg^2$          & $27$--$40$ & $4.6$  & $9.9$\\[1pt]
    \hline
\end{tabular}
\end{center}
\end{table*}

The paper is structured as follows. Section 1 describes our dataset. Section 2 presents our method. Section 3 shows our results on the distribution of galaxies in the MS region and on the evolution of the relation. Section 4 is dedicated to the comparison of our results with those based on the stacking analysis. Section 5 provides a summary of our findings. We  assume  a  $\Lambda$CDM  cosmology  with  $\Omega_M=0.3$, $\Omega_{\Lambda}=0.7$ and $H_0=70$ \textit{km/s/Mpc}, and a Chabrier IMF throughout the paper.

\section{Data and sample selection}

We base the following analysis on two different samples. The GOODS$+$CANDELS sample combines the deep GOODS-Herschel and CANDELS-Herschel observations, with {\it{Spitzer}} MIPS and {\it{Herschel}} PACS coverage over a total area of 760 arcmin$^2$. The COSMOS-PEP sample provides, instead, shallower {\it{Spitzer}} MIPS and {\it{Herschel}} PACS observations over an area of 2 deg$^2$. The GOODS$+$CANDELS sample allows sampling of large part of the MS region with {\it{Herschel}} PACS data at high redshift. The COSMOS-PEP sample provides enough statistics to sample with PACS at any redshift the brightest and rarest infrared systems as the most massive SFGs and SBs.  The stellar masses in the different fields are derived via spectral energy distribution (SED) fitting technique based on different IMFs. They are all converted to a Chabrier IMF for consistency, when needed. Despite the different codes, best fitting procedures and metallicity grids used to obtain the stellar masses in different fields, the estimates are consistent within a scatter of 0.15 dex \citep{2013MNRAS.434.3089Z}. The details of the different datasets are given in the next subsections.

\subsection{GOODS$+$CANDELS dataset}

\subsubsection{The GOODS fields}
For GOODS-S, multiwavelength data (from UV to 24$\mu$m), spectroscopic and photometric redshifts, and stellar masses are provided in the \cite{2015ApJ...801...97S} catalog. This dataset is complemented by PACS data at 70, 100, and 160 $\mu$m, taken from the PEP and GOODS-Herschel catalogs \citep{2013A&A...553A.132M}. The PACS sources are extracted using the FIDEL 24 $\mu$m sources \citep{2007AAS...211.5216D} as prior, down to a $3\sigma$ flux limit of 24 $\mu$Jy. Thus, no additional matching is required between the multiwavelength and the far-infrared catalogs. The fluxes are extracted in all cases via PSF photometry \citep[see for details][]{2009A&A...496...57M,2013A&A...553A.132M}. In the common region, the PEP and GOODS-Herschel dataset reach a 3$\sigma$ level at 0.8 and 2.4 mJy at 100 and 160 $\mu$m, respectively. The GOODS-N multiwavelength data with spectroscopic and photometric redshifts and stellar masses is provided by the PEP catalog \citep{2010A&A...518L..30B}. Similarly to GOODS-S, this dataset is complemented by PACS data at 100 and 160 $\mu$m by the combined PEP and GOODS-Herschel catalogs \citep{2013A&A...553A.132M} and by MIPS FIDEL data at 24 $\mu$m, down to a 3$\sigma$ level of 1.1 mJy, 2.7 mJy and 25 $\mu$Jy, respectively (see also Table 1).

\subsubsection{CANDELS COSMOS and UDS fields}

The CANDELS-\textit{Herschel} survey provides {\it{Herschel}} PACS observations of the UDS field and of an area of 208 arcim$^2$ of the COSMOS region. In COSMOS, in particular, the CANDELS-Herschel and the PEP data have been combined to provide a deeper Herschel coverage to reach the same flux limit of the CANDELS-UDS PACS images (see Table 1 and Inami et al. in prep.). In both fields the source extraction is performed with the same procedure as in the GOODS maps \citep[][Inami et al. in prep.]{2013A&A...553A.132M}.
Spectroscopic and photometric redshifts and stellar masses are taken from \cite{2016ApJS..224...24L} and \cite{2015ApJ...801...97S} for the COSMOS and the UDS area, respectively. Since the MIPS priors used in the PACS source extraction come directly from the multiwavelength catalogs, no cross-matching has to be performed.

The CANDELS MIPS observations of UDS and COSMOS are slightly shallower than in the GOODS fields (see Table 1). Thus, we combine all galaxies in a single sample down to the shallowest flux limit of 40 $\mu$Jy. Between 20 and 40 $\mu$Jy only the GOODS-Herschel $S+N$ sample is considered after applying a volume weight. Throughout the paper we refer to this sample as the GOODS$+$CANDELS sample. 

\subsection{The COSMOS-PEP dataset}
The COSMOS-PEP dataset is provided by the combination of the COSMOS MIPS catalog of \cite{2009ApJ...703..222L} down to a flux limit of 40 $\mu$Jy at the $3\sigma$ level, and the {\it{Herschel}} PACS 100 and 160 $\mu$m catalog as part of the PEP survey \citep{2010ApJ...712.1287L}, down to a flux limit of 4 and 7 mJ at the $3\sigma$ level, respectively, over a 2 deg$^2$ area. This dataset has been matched with UV, optical and near-infrared photometry in the $COSMOS2015^{24}$ catalog \citep{2016ApJS..224...24L}, which contains reliable photometric redshifts and stellar masses for more than half a million objects over the same area. In this sample, 90\% of the 24 $\mu$m-selected sources are securely matched to their $K_s$ band counterpart using the WIRCAM COSMOS map of \cite{2010ApJ...708..202M}, assuming a matching radius of 2". Counterparts to another 5\% of the sample are found using the IRAC-3.6 $\mu$m COSMOS catalog of Sanders et al. (2007), while the rest of the 24$\mu$m sources remain unidentified at shorter wavelengths. 
The source extraction at 24 $\mu$m is performed using the IRACS 3.6 $\mu$m detected source position as prior \citep[e.g.][]{2009A&A...496...57M}. The same exercise is performed for the PACS source extraction of the PEP survey \citep{2011A&A...532A..90L} by using the 24 $\mu$m source position \citep{2013A&A...553A.132M}. The fluxes are extracted in all cases via PSF photometry \citep[see][]{2009ApJ...703..222L,2013A&A...553A.132M}. Throughout the paper we refer to this sample as the COSMOS-PEP sample.

\subsection{The UV+IR SFR at high redshift}

Far-infrared  (FIR) and ultraviolet (UV) emissions are the two key components for accurately measuring the SFR of an object. Part of the UV emission originated from the young star population is absorbed by dust and re-processed at infrared wavelengths. Such emission alone can provide a measure of the SFR only if corrected for this absorption. However, the measure of the dust attenuation is still very uncertain because of the degeneracies between age and reddening, the assumption on galaxy metallicity and SF histories and the parametrization of the extinction curve \citep[e.g.][]{1999ApJ...521...64M,2009ApJ...703..517D,2017MNRAS.466..861D,2017MNRAS.467.1360B}. As a result, the combination of IR and UV emission is considered the most accurate way to determine the galaxy SFR \citep[see][and references therein]{2014ARA&A..52..373L}. Here we describe how the two contributions are estimated for each galaxy. 

\begin{figure}
	\includegraphics[width=\columnwidth]{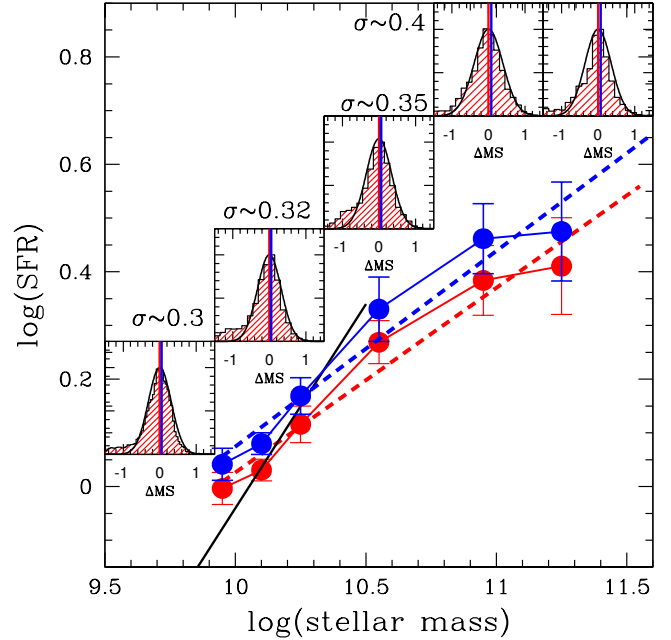}
	\label{riassunto}
\caption{Location of the local MS based on WISE 22 $\mu$m derived SFRs {\protect\citep{2019MNRAS.483.3213P}}. The red and blue points show the location of the {\it{median}} and {\it{mean}} SFR, respectively, of the best fit log-normal distribution in the local Universe. The figure shows also the best linear regression for the median MS (dashed red line) and for the mean MS (blue dashed line) and the median MS relation of {\protect\cite{2015ApJ...801L..29R}} below $10^{10.5}$ $M_{\odot}$ (black solid line). For each stellar mass interval we show also the distribution of the residuals around the MS. The residual $\Delta$MS is defined as log(SFR$_{gal}$)-log(SFR$_{MS}$). The black solid curve in the small panels shows the best fit log-normal relation obtained in {\protect\cite{2019MNRAS.483.3213P}}. The red and blue solid vertical lines in each panel indicate the median and mean SFRs, respectively. For each panel we indicate also the value of the dispersion of the log-normal component $\sigma$, which increases as a function of stellar mass.}
\end{figure}

For galaxies observed either by {\it{Herschel}} PACS or by {\it{Spitzer}} MIPS, we compute the IR luminosity, $L_{IR}$, as follows. We use all the available mid- and FIR data-points to compute $L_{IR}$ by integrating the best fit SED template in the range 8-1000\,$\mu$m. To this aim we use two different sets of templates to check the model dependence. We use the MS and SB templates of \cite{2011A&A...533A.119E} and the set of \cite{2012ApJ...758L...9M}. All templates provide infrared luminosities consistent with each other with a rms of 0.05 dex (see Appendix A). As an alternative method, we compute LIR by fitting the full SED with hyperzspec, using the templates of \cite{2003MNRAS.344.1000B} for near-UV to near-IR wavelengths, and the ones of \cite{2007ApJ...663...81P} for mid-IR and FIR wavelengths.  For sources with both MIPS and PACS detections, all methods provide $L_{IR}$ consistent with each other with a rms ranging from 0.08 to 0.15 dex (see Appendix A).

For undetected PACS sources, we rely only on the 24 $\mu$m flux. Because without FIR information it is not possible to determine the best fit between the MS or SB templates of \cite{2011A&A...533A.119E}, we adopt a MS template. To test the reliability of the 24 $\mu$m based $L_{IR}$, we compare it with $L_{IR}$ derived from PACS fluxes for the PACS detected sources. We assume as reference the MS location of \cite{2014ApJ...795..104W} at different redshifts. The $L_{IR}$ based on the 24 $\mu$m flux is consistent with the PACS derived $L_{IR}$ for galaxies below and on the MS. The mean ratio of the two estimates is 1 (Appendix B, Fig. A2). Above the MS location, the larger the distance from the MS, the larger the underestimation of the $L_{IR}$ based on the 24 $\mu$m point with respect to the PACS based estimates. The effect is redshift dependent because it is related to the PAH emissions around 8 and 12 $\mu$m rest-frame, which enter the MIPS 24 $\mu$m filter above $z\sim0.5$. In addition to the different dust temperatures, the need of a MS and SB templates is due to the different equivalent width of the PAH emission, which decreases at increasing distance from the MS \citep{2011A&A...533A.119E, 2012ApJ...758L...9M}. Thus, the use of the MS template leads to an underestimation of the $L_{IR}$ above the MS, while the use of the SB template leads to overestimated $L_{IR}$ for galaxies on or below the MS. The same bias is observed when using the templates of \cite{2012ApJ...758L...9M}, which are built on the basis of the same dataset used in \cite{2011A&A...533A.119E}. Thus, without the key information provided by FIR data, the $L_{IR}$ based on the 24 $\mu$m flux are reliable only when estimated for galaxies below or on the MS with MS galaxy infrared templates. For galaxies at SFR larger than the MS location, the use of PACS data is necessary.

Once estimated, the IR luminosity is then converted into dust-reprocessed ${\rm SFR}$ using the Kennicutt (1998) formula:
\begin{equation} 
    {SFR}_{IR} = 1.05\times10^{-10}\,L_{IR}\,[L_\odot]\,.
\end{equation}

The rest-frame UV luminosity is estimated at 1500 \AA \hspace{0.05cm} by using the SED fitting technique with {\it{LePhare}} \citep{1999MNRAS.310..540A,2006A&A...457..841I}. The UV luminosity is then converted into SFR uncorrected for dust attenuation using the formula from \cite{2004ApJ...617..746D}, i.e.,
\begin{equation}
    {SFR}_{UV} = 2.17\times10^{-10}\,L_{UV}\,[L_\odot]\,.
\end{equation}

The total ${\rm SFR}$ is finally computed as the sum of ${\rm SFR}_{\rm UV}$ and ${\rm SFR}_{\rm IR}$. The two relations above are derived assuming a Salpeter IMF and are then corrected to a Chabrier IMF.

\begin{table*}
\begin{center}
\begin{tabular}{lccccc}
    \hline
    \hline \\[-2.5mm]
    redshift & $10^{10-10.2}$ $M_{\odot}$ & $10^{10.2-10.5}$ $M_{\odot}$ & $10^{10.5-10.8}$ $M_{\odot}$ & $10^{10.8-11.2}$ $M_{\odot}$ & $ >10^{11.2}$ $M_{\odot}$ \\
  \hline
    \hline \\[-2.5mm]
$0 < z < 0.3$ & \color{magenta} 570 & \color{magenta} 241 &\color{magenta} 188 & \color{magenta}113 & \color{magenta} 35\\
${\chi}^2$ & 1.2 & 1.3 & 1.1 & 1.45 & 2 \\
$P_{KS}$ & 99\% & 98\% & 98\% & 95\% & 75\% \\
 \hline
$0.3 < z < 0.5$ & \color{magenta}1138 & \color{magenta} 819 &\color{magenta}  612 & \color{magenta} 392 & \color{magenta}90    \\
${\chi}^2$ & 0.9 & 0.95 & 1.1 & 1.15 & 2.5 \\
$P_{KS}$ & 99\% & 98\% & 98\% & 99\% & 70\% \\
 \hline
$0.5 < z < 0.8$ & \color{blue} 346 & \color{blue}  203 & \color{blue} 139 & \color{blue} 81 &\color{magenta} 193  \\
${\chi}^2$ & 1.1 & 1.05 & 0.85 & 1.05 & 1.7 \\
$P_{KS}$ & 97\% & 99\% & 98\% & 99\% & 80\% \\
 \hline
$0.8 < z < 1.2$ & \color{blue} 354 & \color{blue} 233 & \color{blue} 225 & \color{blue} 188 &\color{magenta} 302  \\
${\chi}^2$ & 1.4 & 1.2 & 1.2 & 1.2 & 1.8 \\
$P_{KS}$ & 95\% & 97\% & 97\% & 98\% & 75\% \\
 \hline
$1.2 < z < 1.6$ & \color{blue} 201 & \color{blue} 185 & \color{blue} 152 & \color{blue} 95 &\color{magenta} 214 \\
${\chi}^2$ & 1.35 & 1.2 & 1.3 & 1.3 & 1.3 \\
$P_{KS}$ & 95\% & 96\% & 97\% & 97\% & 93\% \\
 \hline
$1.6 < z < 2.2$ &  & \color{blue} 139 & \color{blue} 152 & \color{blue} 114 &\color{magenta} 103 \\
${\chi}^2$ &  & 1.3 & 1.45 & 1.3 & 1.4 \\
$P_{KS}$ &  & 95\% & 96\% & 96\% & 94\% \\
 \hline
$2.2 < z < 2.5$ &  & \color{blue} 160 & \color{blue} 150 & \color{blue} 131 & \color{magenta}194 \\ 
${\chi}^2$ &  & 1.4 & 1.3 & 1.35 & 1.8 \\
$P_{KS}$ &  & 95\% & 96\% & 95\% & 83\% \\
 \hline
   \hline
\end{tabular}
\end{center}
\caption{The table lists the number of galaxies, the reduced $\chi^2$ and the KS probability that local and higher redshift $\Delta{MS}$ distribution are drawn from the same parent distribution, above the 80\% completeness level. The galaxy number is indicated in magenta when the data are taken from the COSMOS PEP sample, and in blue when analysis is based on the GOODS$+$CANDELS sample.}
\end{table*}

\subsection{The adopted strategy}

Rather than directly measuring the location of the MS and inferring its evolution, we start from the null hypothesis that the MS is evolving across cosmic time only in normalization and not in shape and slope. Namely, we assume as null hypothesis that the MS location at $z>0$ is given by:
\begin{equation}
SFR= MS_{0}*{\gamma}(z)
\label{nn}
\end{equation}
where \textit{MS$_{0}$} is the local relation, and \textit{$\gamma(z)$} is the evolution of the normalization. We define as residual around the MS the value $\Delta${MS}=log(SFR$_{gal}$)-log(SFR$_{MS}$(z)), where SFR$_{gal}$ is the galaxy SFR and SFR$_{MS}(z)$ is the value of SFR on the MS at the same stellar mass and redshift. We assume as part of the null hypothesis that the distribution of the residuals per mass bin is the same at all redshifts and consistent with the distribution observed in the local Universe, as suggested by \cite{2015A&A...579A...2I} up to $z\sim 1.4$.

As local MS we use the relation obtained at $z < 0.085$ from WISE 22 $\mu$m SFRs by \citep[][see also Fig. \ref{riassunto}]{2019MNRAS.483.3213P}. This is consistent with the one based on other SFR indicators, such as the $H{\alpha}$ emission and the infrared luminosity based on {\it{Herschel}} PACS and SPIRE data \citep[see][for an extensive comparison]{2019MNRAS.483.3213P}. As shown in the inner panels of Fig. \ref{riassunto}, galaxies are distributed with a log-normal distribution in the MS region. The location of the MS is defined as the location of the peak of the log-normal distribution, as in \cite{2015ApJ...801L..29R}, which is the median SFR of the distribution. We define as upper and lower envelope of the MS, the regions above and below the peak of the distribution, respectively.

Due to the bias identified in the previous section, some caution must be taken in the use of the 24 $\mu$m based $L_{IR}$ in the study of the SB region and of the scatter of the MS, as already pointed out in \cite{2015A&A...575A..74S}. To overcome this problem, we make sure to sample with {\it{Herschel}} PACS data the whole MS upper envelope at least 1$\sigma$ above the peak of the relation (where the $\sigma$ is given by the dispersion of the log-normal relation shown in Fig. \ref{riassunto}). Below this limit we sample the peak and the lower envelope of the relation with {\it{Spitzer}} MIPS 24 $\mu$m data. 

With this restriction, the COSMOS-PEP dataset allows to study the MS only at $z < 0.7$ or at very high masses ($> 10^{10.8} M_{\odot}$) at higher redshift. In all other cases, we use the GOODS$+$CANDELS dataset.

X-ray detected AGN account for 10\% of the sample \citep{2011A&A...532A.145P,2012A&A...537A..58P}. Most of them (65\%) are observed with Herschel PACS at FIR wavelengths, where the AGN contribution is negligible with respect to the star formation contribution \citep{2014ARA&A..52..373L}. The rest is observed at 24 $\mu$m data only, where the AGN contribution could be more significant. This subsample, which account for a marginal fraction, is removed from the final galaxy sample.

\begin{figure*}
\includegraphics[width=17cm,height=18cm]{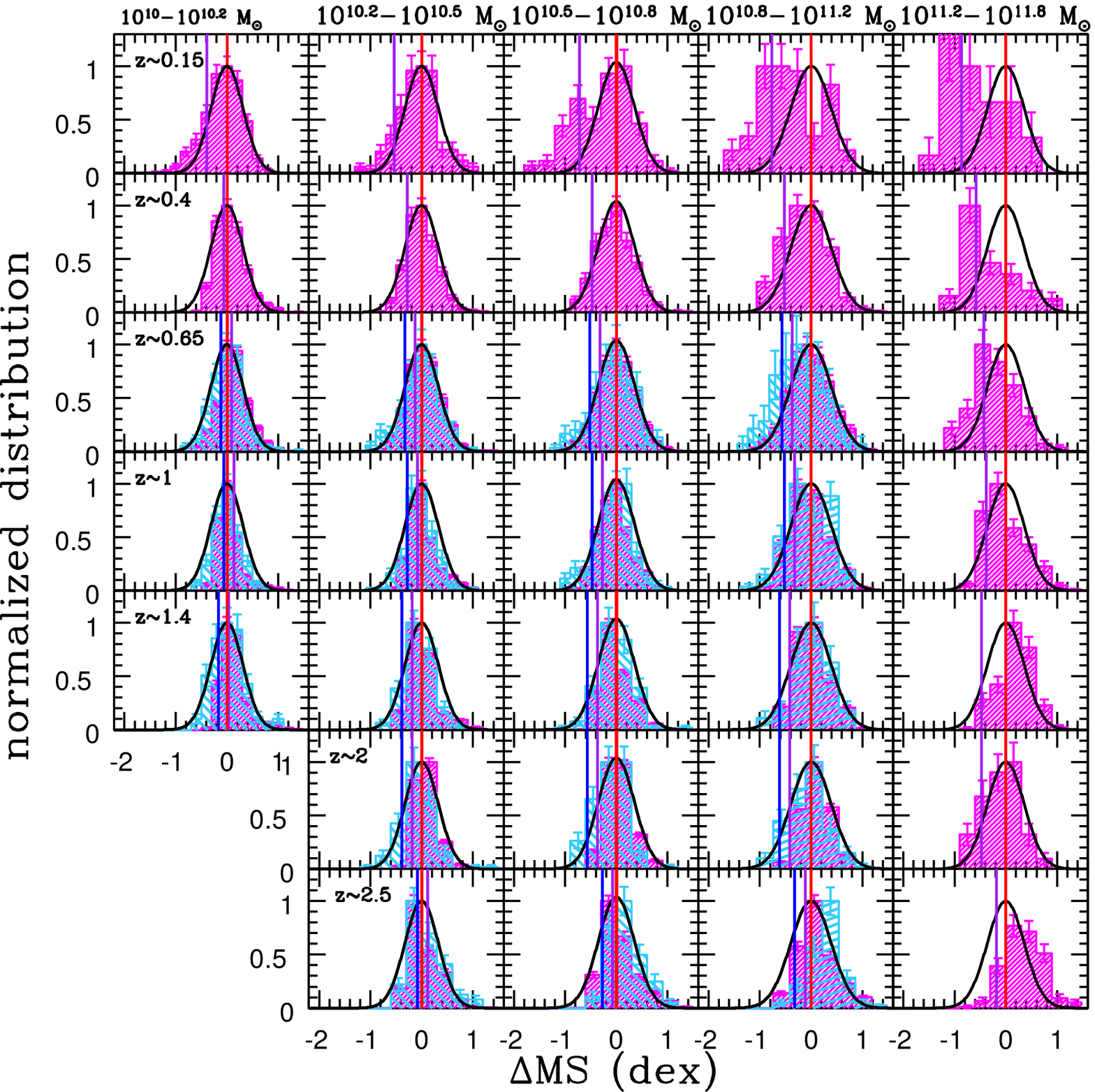}
\caption{Evolution of the distribution of galaxies around the MS as a function of redshift in several stellar mass bin. The redshift is increasing in the vertical direction from the top to the bottom, while the stellar mass is increasing along the horizontal direction from left to right. The magenta and blue histograms in each panel represent the distribution of the COSMOS and of the CANDELS+GOODS galaxies around the MS, respectively. The black curve in each panel is the local best fit distribution obtained at z$\sim$ 0 by \protect\cite{2019MNRAS.483.3213P} and shown in Fig. \ref{riassunto}. The vertical red line shows the $\Delta${MS}=0 point corresponding to the local MS location shifted rigidly by the best fit value of the normalization $\gamma(z)$. The vertical purple and blue lines show the 80\% completeness limit of the COSMOS and GOODS$+$CANDELS sample, respectively.}
\label{ms_ev}
\end{figure*}

\begin{figure}
\includegraphics[width=\columnwidth]{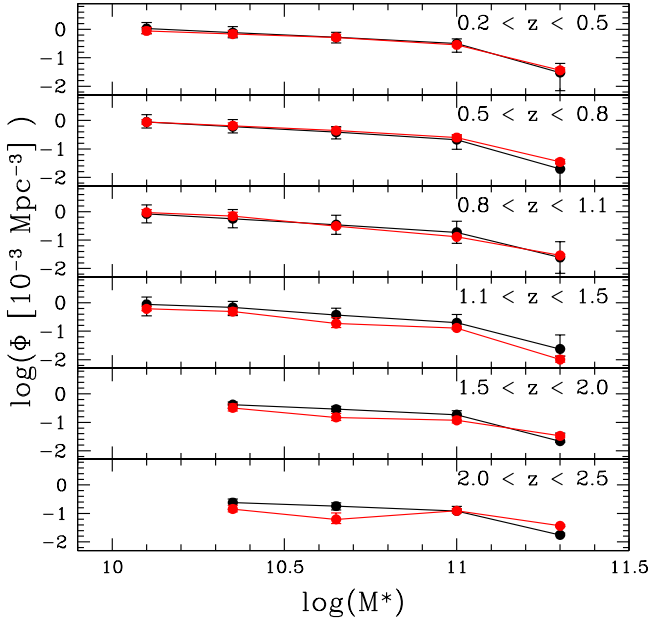}
\caption{Comparison of the comoving number density of SF galaxies derived from the SF galaxy stellar mass function of \citet[][black points and line]{2013A&A...556A..55I}, versus the comoving number density of MS galaxies (red points and line) in several stellar mass bins and redshifts.}
\label{norma_ilbert}
\end{figure}

\begin{figure}
\includegraphics[width=\columnwidth]{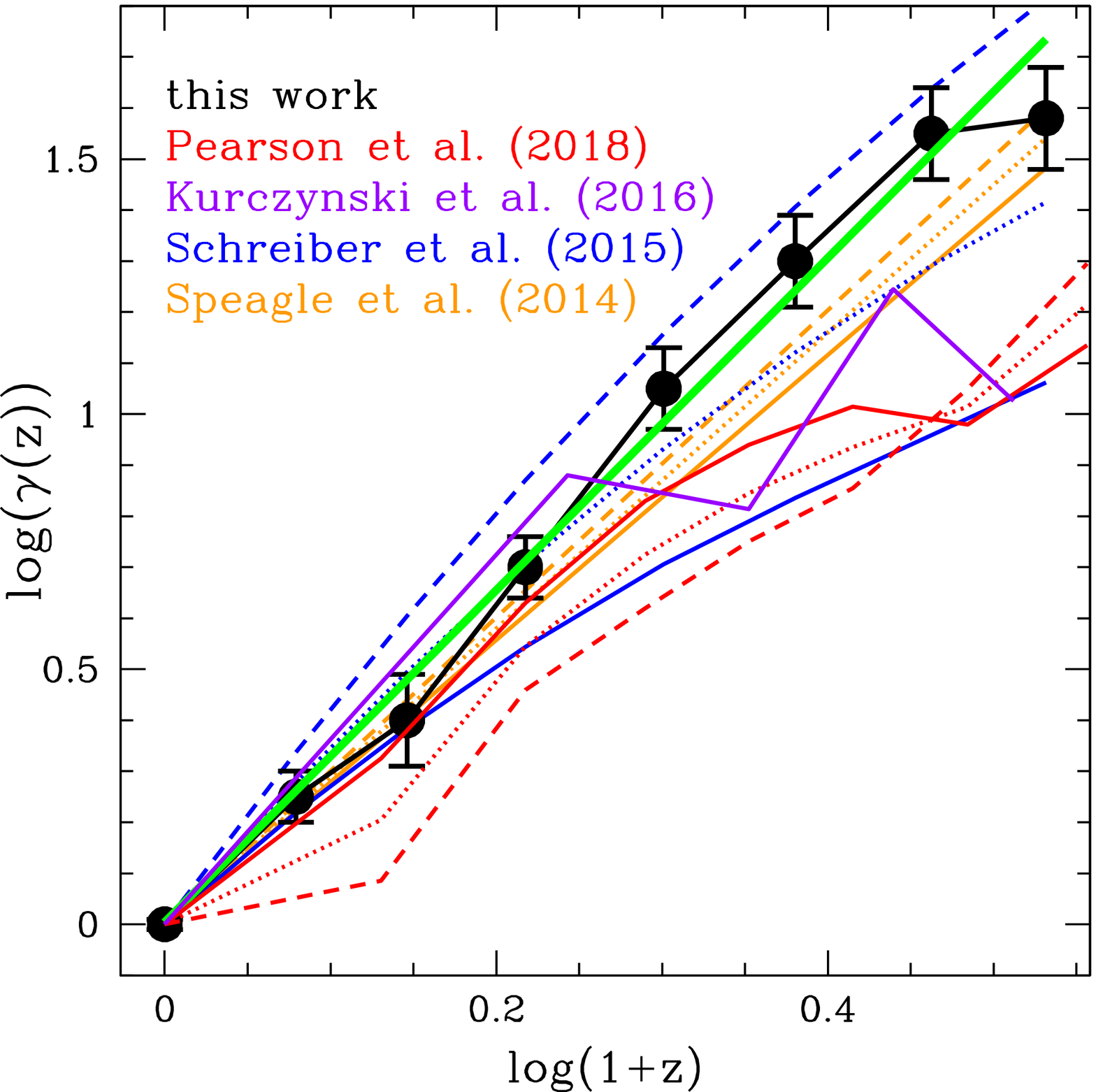}
\label{norma}
\caption{Evolution of the normalization $\gamma(z)$ of the MS as a function of redshift. The result of our method is shown by the black points. The green line indicates our best fit of the form $\gamma(z)\propto (1+z)^{\alpha}$. We report also the evolution based on previous results in the literature at three stellar mass values: $10^{10}$ $M_{\odot}$ (solid lines), $10^{10.5}$ $M_{\odot}$ (dashed lines) and $10^{11}$ $M_{\odot}$ (dotted lines). The error bars of the individual lines are not drawn for sake of simplicity. For marginally evolving MS slopes, as in the case of \protect\cite{2014ApJS..214...15S} and Pearson et al. (2018), the three lines are consistent within 1$\sigma$. For evolving MS slopes as in the case of \protect\cite{2015A&A...575A..74S}, the three lines are not consistent. We do not report the evolution obtained by \protect\cite{2014ApJ...795..104W} and \protect\cite{2016ApJ...817..118T} for sake of simplicity, because they are consistent with \protect\cite{2015A&A...575A..74S}. }
\label{norma}
\end{figure}

\section{The evolution of the MS}

\subsection{The evolution of the MS location at the high mass end}

In order to check the validity of the null hypothesis, we divide our IR selected galaxy sample in 7 redshift bins from $z=0.3$ to $z=2.5$. The number of galaxies per mass and redshift bin is given in Table 2.

If the null hypothesis is valid, the MS at higher redshifts is just a rescaled version of the local relation with a higher normalization $\gamma(z)$ (see Eq. \ref{nn}). To check this hypothesis, we shift rigidly the best fit distributions of the local MS residuals to higher values of SFR to match the distributions of galaxies at higher redshift in any stellar mass bin. We exclude from the best fitting procedure the lowest and the highest stellar mass bins. In the former case the exclusion is beacuse the completeness is not sufficient to properly sample the peak and the lower envelope of the distribution in the highest redshift bins. In the latter case, the statistics is poor and the uncertainty of the local distribution is large \citep{2019MNRAS.483.3213P}. The best fit of $\gamma(z)$ at all redshifts is, thus, 
measured as the rigid shift providing the best simultaneous match over a range of 1 order of magnitude in stellar mass, from $10^{10.2}$ to $10^{11.2}$ $M_{\odot}$. Namely, we measure the $\chi^2$ separately in the three stellar mass bins and we define as best fit the one givin the lowest mean reduced  $\chi^2$. The best fit local distribution and the observed high redshift distribution are normalized to a common $\Delta$MS region at $0 <$ $\Delta$MS $< 0.5$ before being matched. This limits us to fit purely $\gamma(z)$.

We check the validity of the null hypothesis by assessing the $\chi^2$ value between the observed binned residual distribution and the $z\sim0$ best fit residual distribution, per mass bin. In addition we perform a KS test on the unbinned data. The values of the reduced $\chi^2$ and KS statistics are listed in Table 2 for each redshift and stellar mass bin. We consider the null hypothesis valid when the reduced $\chi^2$ is lower than 1.5 and the KS probability of being drawn from the same parent distribution is higher than 95\%.

The results of such experiment are shown in Fig. \ref{ms_ev}. The solid black curves in each panel shows the best fit of the local log-normal distribution of \cite{2019MNRAS.483.3213P} in different mass bins, as shown in the small inner panels of Fig. \ref{riassunto}. The histograms show the residual distributions of the COSMOS-PEP (magenta) and GOODS$+$CANDELS (cyan) samples in different redshift and mass bins. The COSMOS-PEP data are used for the $\chi^2$ and KS tests only at $z<0.7$ and at M$_*$ $>10^{10.8}$ $M_{\odot}$ at higher redshift. The histograms are corrected for incompleteness by applying a $1/V_{max}$ correction, computed exploiting the template used to derive $L_{IR}$. This procedure implicitly assumes no strong evolution of the number density of the population in the small redshift windows probed here. The purple and blue vertical lines in each panel of Fig. \ref{ms_ev} show the 80\% completeness limit for the COSMOS-PEP and GOODS$+$CANDELS samples, respectively. This is obtained, as for the $1/V_{max}$ correction, by shifting the IR template to higher redshift to estimate the minimum $L_{IR}$ (SFR) corresponding to the 24 $\mu$m 5$\sigma$ limit, which leads to the observation of 80\% of the galaxy population in the given redshift window. The experiment is repeated by estimating $L_{IR}$ from three different sets of IR templates \citep{2007ApJ...663...81P,2011A&A...533A.119E, 2012ApJ...758L...9M}. The histograms are consistent to each other within 1$\sigma$ error bars. Fig. \ref{ms_ev} shows, in particular, the results based on the \cite{2011A&A...533A.119E} templates.

In all considered bins below $z \sim 2.2$, the reduced $\chi^2$ value between the observed \textit{$\Delta$MS} distribution at high redshift and the best fit local log-normal distribution is comprised between 0.85 and 1.45 above the 80\% completeness limit. The KS probability is above $95\%$. Thus, the null hypothesis is considered valid. In the lowest stellar mass bin at $10^{10-10.2}$ $M_{\odot}$, not included in the fitting procedure, the reduced $\chi^2$ is in the range 0.9 and 1.4 with a KS probability above $95\%$. We consider the null hypothesis valid also at this stellar masses. In the bins with the poorest statistics, at $2.2 < z < 2.5$ and at masses above $10^{11.2}$ $M_{\odot}$, the reduced $\chi^2$ value is comprised between 1.3 and 2.5, while the KS probability lower limit is $~70\%$. In this bins we consider the result as inconclusive.

In order to test any effect due to the mid- and far-infrared selection and the completeness of our sample, we compare the number density of MS galaxies with the number density of SFGs estimated through the SF stellar mass function of \cite{2013A&A...556A..55I}. In this work, SFGs are selected in the COSMOS field according to their absolute ($M_{NUV}$- $M_R)-(M_R$-$M_J)$ colors. For this exercise, the number density of MS galaxies is estimated by doubling the number density of galaxies observed at $\Delta{MS} > 0$ at any redshift and stellar mass bin. This is done because the SFR at $\Delta{MS}= 0$ is the median SFR of the log-normal distribution. The error is mainly driven by the uncertainty in the MS normalization, $\gamma(z)$, than the Poissonian error. The integration of the double Schechter function of \cite{2013A&A...556A..55I} in different stellar mass bins, provides the number density of SFGs at different redshifts. The error is estimated by marginalizing over the uncertainties of the double Schechter best fit parameters. Fig. \ref{norma_ilbert} shows the results of the comparison. The SF (black line and points) and MS (red line and points) galaxy number densities are remarkably in agreement within 1-1.5$\sigma$. This suggests that our results are not affected by strong incompleteness or selection effects.

\begin{table}
\begin{center}
\begin{tabular}{ccc}
    \hline
    \hline \\[-2.5mm]
    redshift & $M^*/M_{\odot}$ & log(SFR) \\
  \hline
    \hline \\[-2.5mm]
0.05 & $10^{10-10.2}$ $M_{\odot}$  &  0.030$\pm$0.022 \\
0.05 & $10^{10.2-10.5}$ $M_{\odot}$&  0.115$\pm$ 0.034 \\
0.05 & $10^{10.5-10.8}$ $M_{\odot}$&  0.268$\pm$ 0.041 \\
0.05 & $10^{10.8-11.2}$ $M_{\odot}$&  0.383$\pm$ 0.065 \\
0.05 & $ >10^{11.2}$ $M_{\odot}$   &  0.410$\pm$0.093 \\
 \hline
   \hline
\end{tabular}
\begin{tabular}{cc}
    redshift & log($\gamma$(z)) \\
  \hline
    \hline \\[-2.5mm]

$0 < z < 0.3$   &0.25$\pm$0.05 \\
$0.3 < z < 0.5$ &0.40$\pm$0.09 \\
$0.5 < z < 0.8$ &0.75$\pm$0.06 \\
$0.8 < z < 1.2$ &1.10$\pm$0.08 \\
$1.2 < z < 1.6$ &1.35$\pm$0.09\\
$1.6 < z < 2.2$ &1.55$\pm$0.09\\
$2.2 < z < 2.5$ &1.58$\pm$0.11 \\
 \hline
   \hline
\end{tabular}
\end{center}
\label{gamma1}
\caption{The upper table lists the value of the MS location at $z\sim0$ ($MS_0$) of Popesso et al. (2019), as in of Eq. \ref{nn}. The bottom table lists the evolution of the normalization $log(\gamma(z))$  with respect to the local relation, as a function of redshift, as in Eq. \ref{nn}.}
\end{table}

\begin{figure*}
\includegraphics[width=17cm, height=13cm]{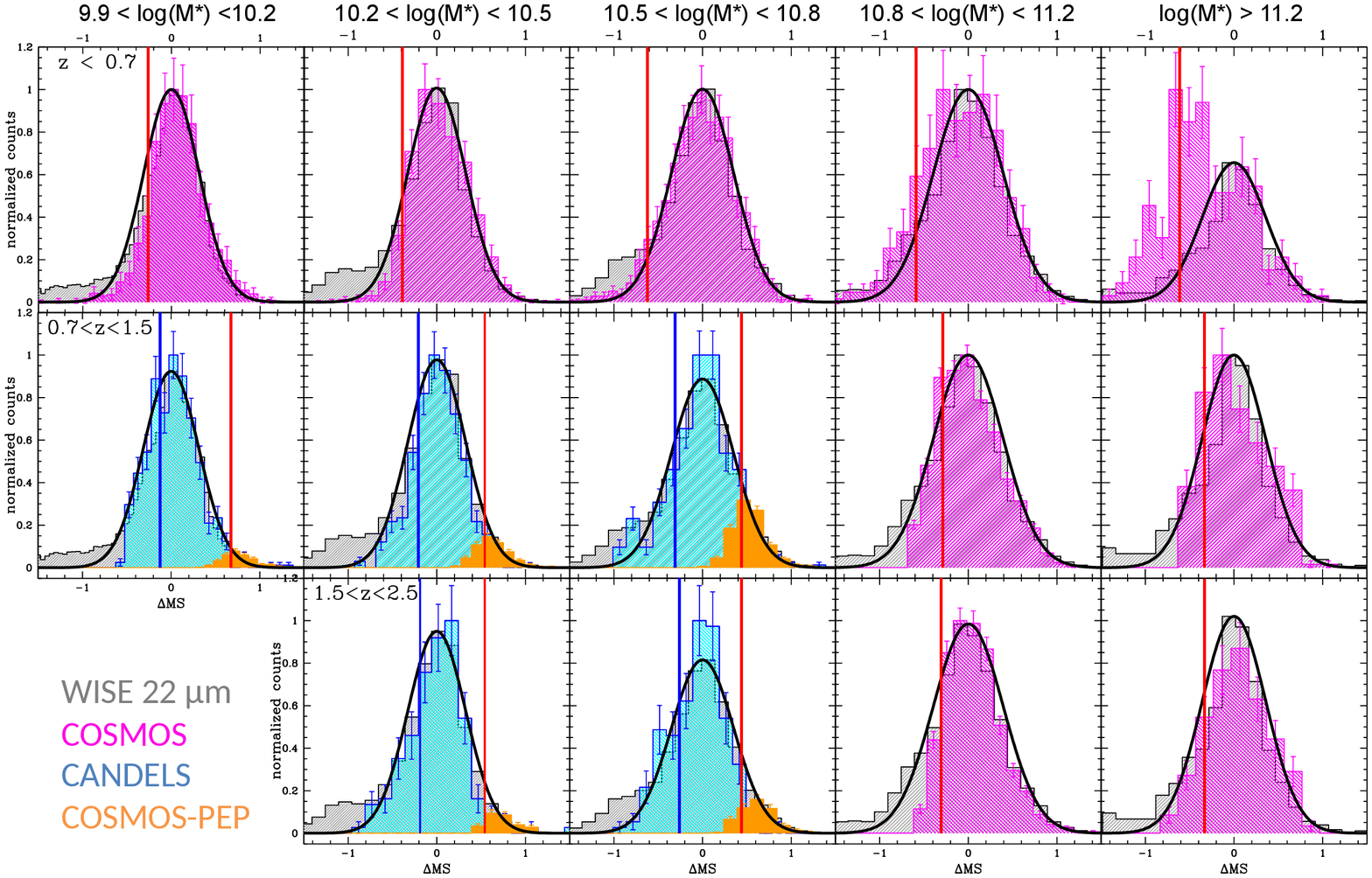}
\caption{Evolution of the SFR distribution in the MS region as a function of stellar mass in three redshift bins at $z < 0.7$ (upper panels), $0.7 < z < 1.5$ (central panels) and $1.5 < z < 2.5$ (bottom panels). The redshift is increasing in the vertical direction, while the stellar mass is increasing along the horizontal direction. The gray shaded histogram shows the distribution of local galaxies around the local MS as shown in Fig. \ref{riassunto}. The magenta and blue histograms in each panel show the distributions of the COSMOS and of the CANDELS+GOODS galaxies, respectively. The orange histogram in the second and third rows of panels shows the distribution of the COSMOS-PEP subsample limited to PACS detections. The black curve in each panel is the best fit local relation obtained by fitting the distribution of WISE 22$\mu$ based SFRs MS upper envelope in \protect\cite{2019MNRAS.483.3213P}. The local best fit is scaled to match the COSMOS and CANDELS+GOODS distributions in a common $\Delta{MS}$ region at $0 < \Delta{MS} < 0.5$. For simplicity all distributions are then re-normalized to the observed peak. The vertical red and blue lines show the 80\% completeness limit in the COSMOS-PEP and CANDELS+GOODS sample, respectively. }
\label{combo}
\end{figure*}

\subsection{The evolution of the MS normalization}

In Eq. \ref{nn}, the factor $\gamma(z)$ expresses the evolution of the MS normalization with respect to the local relation. The values of $\gamma(z)$ as a function of redshift are shown in Fig. \ref{norma} (black points). The best fit (green line) is a power law of the form: 
\begin{equation}
\gamma(z)=1.01\pm{0.03}\times(1+z)^{3.21\pm{0.19}}
\label{eq_z}
\end{equation}
The values of $\gamma(z)$ as a function of redshift are reported in Table 3. We also fit $\gamma(z)$ by shifting separately the local distribution of the individual stellar mass bins. However, we point out that the error bars of the normalization in this case are 2 to 3 times larger than the error of the constant normalization. This is due to the limited statistics of the individual bins, which prevents an accurate localization of the peak of the $\Delta{MS}$ distribution. For this reason, all results are statistically consistent (within 1$\sigma$) with each other and with the constant normalization per stellar mass.  

Fig. \ref{norma} compares the evolution of $\gamma(z)$ obtained here with previous results in the literature. The redshift dependence of $\gamma(z)$ is in all cases best fitted with a power law, $\propto(1+z)^{\alpha}$, with $\alpha$ varying from 2 to 3.8 \citep[e.g.][]{2009ApJ...698L.116P, 2010MNRAS.405.2279O,2011ApJ...730...61K,2014ApJS..214...15S, 2015A&A...575A..74S, 2015A&A...579A...2I, 2016ApJ...820L...1K,2018A&A...615A.146P}. The variation of $\alpha$ depends on the use of different SFR indicators, as pointed out by \cite{2014ApJS..214...15S}, but also on the stellar mass, if the MS is fitted with an evolving slope. Studies which find a marginally evolving slope up to redshift $\sim2.5$ \citep{2014ApJS..214...15S,2016ApJ...820L...1K,2018A&A...615A.146P}, show that $\gamma(z)$ has a marginal mass dependence and it evolves as $\propto(1+z)^{2.6-2.9}$. Conversely, if the MS steepens at higher redshift \citep[e.g.][]{2014ApJ...795..104W,2015A&A...575A..74S,2016ApJ...817..118T}, the exponent ${\alpha}$ exhibits a large excursion as a function of the stellar mass, ranging from $\sim2.6\pm0.1$ at $10^{10}$ $M_{\odot}$ to $\sim3.5\pm0.15$ at $10^{11}$ $M_{\odot}$ \citep[see also][]{2015A&A...579A...2I}. Our estimate is in agreement within 1.5$\sigma$ with \citet[][$\alpha=2.9\pm0.1$ at $10^{10.5}$ $M_{\odot}$]{2014ApJS..214...15S}, and within 1$\sigma$ with previous findings leading to $3 < \alpha < 3.4$ \citep{2010ApJ...714L.118D,2010MNRAS.405.2279O,2011ApJ...730...61K}. However, it is not in agreement with the results of \cite{2016ApJ...820L...1K} and \cite{2018A&A...615A.146P}, who find a much flatter evolution. Both studies find a marginally evolving slope of the MS. However, we point out that \cite{2016ApJ...820L...1K} sample only few tens of galaxies above $10^{10}$ $M_{\odot}$. Thus, their MS is mainly driven by lower mass galaxies. In the case of \cite{2018A&A...615A.146P}, instead, the discrepancy is ascribable to a flat MS, which exhibits an offset of 0.4 dex below the values of \cite{2014ApJS..214...15S} and \cite{2015A&A...575A..74S} at $1 < z < 2.5$, as reported by the authors.

\subsection{Evolution of the MS scatter}

The limited statistics due to the small redshift bins in Fig. \ref{ms_ev} does not allow to accurately check the evolution of the scatter with respect to the local Universe. To test also this aspect, we increase the statistics by collapsing the 7 redshift bins of Fig. \ref{ms_ev} in 3 larger bins by taking into account the evolution of the MS location. Namely, we measure for each galaxy the residual \textit{$\Delta$MS} with respect to the MS location at the galaxy redshift, by shifting the local MS according to Eq. \ref{eq_z}. This is done to remove any evolutionary effect that could bias the shape of the MS in the large redshift bins considered here (see Appendix B and Fig. \ref{ev_bias} for a detailed discussion).  Each galaxy is weighted according to the $1/V_{max}$ method, as described above. As for the previous test, we use the COSMOS-PEP sample at $z < 0.7$ and at the high stellar masses, and the GOODS$+$CANDELS data elsewhere.

The results of this analysis are shown in Fig. \ref{combo}. The vertical blue and red lines in each panel show the 80\% completeness limit in each redshift bin as provided by the $1/V_{max}$ method. Below this limit the histograms are fully dominated by the completeness correction and likely underestimate the actual distribution. Thus, when performing the $\chi^2$ or the KS test, we rely only on the part of the histograms above these thresholds. We point out that, in the majority of the cases, this limit prevents studying the shape of the MS lower envelope. Thus, most of the analysis focuses on the log-normal component of the MS and on constraining, in particular, its upper envelope. As done in Fig. \ref{ms_ev}, the high redshift and the local best fit log-normal distributions are normalized to a common region between the MS peak and $+$0.5 dex towards the upper envelope. This region is chosen to be sampled at all redshifts in all stellar mass bins and to exclude the tail of the MS at high SFR, which might deviate from a single log-normal component, as suggested in \cite{2012ApJ...747L..31S}. We point out that no fitting is performed in this case. The local relation is simply normalized consistently with the high redshift distribution.

\begin{figure}
\includegraphics[width=\columnwidth]{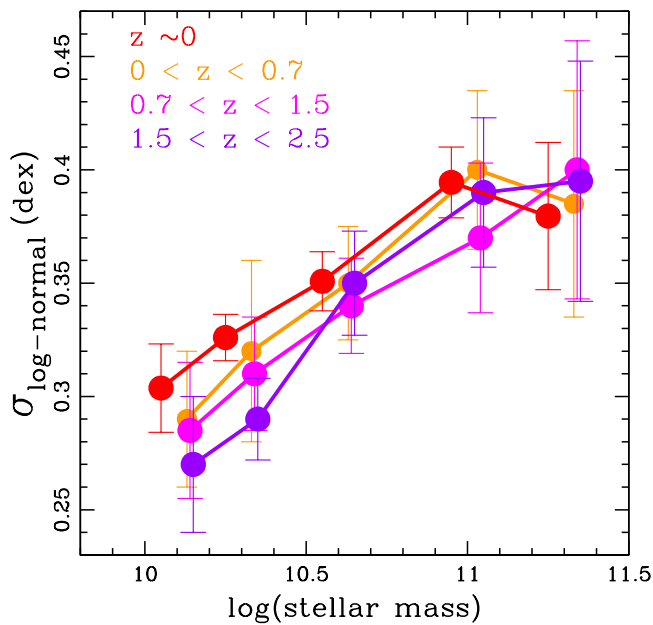}
\caption{Evolution of the scatter of the MS at different redshifts. The scatter is measured as the dispersion of the log-normal component.}
\label{scatter_ev}
\end{figure}

At all redshift and stellar mass bins below $~10^{11.2} M_{\odot}$, the KS and the $\chi^2$ tests reveal a remarkably good agreement down to the 80\% completeness level, between the observed high redshift and the local distributions. Only in the highest stellar mass bin, above $10^{11.2} M_{\odot}$, the KS and $\chi^2$ tests are inconclusive at any redshift despite the larger statistics. This result suggests that, as in the local Universe, the scatter of the MS increases towards higher masses. To further check this result we also fit the high redshift distributions independently in any stellar mass bin with a log-normal component above the 80\% completeness level. In all cases we obtain a best fit consistent within 1$\sigma$ with the local best fit. We take the dispersion of the individual log-normal component per stellar mass bin as a measure of the scatter around the MS. Fig. \ref{scatter_ev} shows the dependence of the scatter as a function of stellar mass from $z\sim 0$ to $z\sim2.5$.  As in the local Universe, the scatter of the relation is increasing as a function of stellar mass from $\sim 0.28\pm0.03$ dex at $10^{10}$ $M_{\odot}$ to $\sim0.39\pm0.03$ at $10^{11}$ $M_{\odot}$. Due to the lower statistics of the high redshift distributions with respect to the local one, the error bars are larger and the trend is significant only at the 2$\sigma$ level. The reported values are for the observed scatter. By taking into account the measurement uncertainties of the SFR and stellar mass estimates, these should be reduced by 25\% to obtain the intrinsic scatter.

The scatter around the MS is found to vary in the literature from 0.2 to 0.35 dex \citep[see][]{2014ApJS..214...15S,2007ApJ...670..156D,2010ApJ...714L.118D,2010ApJ...714.1740M,2012ApJ...754L..29W, 2015A&A...575A..74S}. Previous studies agree in finding a relatively non evolving scatter as a function of time \citep{2016ApJ...820L...1K,2018A&A...615A.146P}.  \cite{2015A&A...579A...2I} explore the mass dependence of the scatter and find results consistent with those presented here up to z$\sim$1.4. All other works in the literature do not explore or do not find any stellar mass dependence.



\subsection{The SB region}

We use the large statistics shown in Fig. \ref{combo} to test the SB region of the SFR distribution. \cite{2012ApJ...747L..31S} suggest that an excess of SBs is required in order to justify the observation of the double power law infrared luminosity function \citep[e.g.][]{2009A&A...496...57M,2013A&A...553A.132M}. An indication of an excess of SBs  at $1.5 < z < 2.5$ is observed in \cite{2011ApJ...739L..40R} in the COSMOS field, after combining deep UV based and IR derived SFRs. \cite{2015A&A...575A..74S} provide similar evidence in a larger redshift window, by combining {\it{Spitzer}} MIPS 24 $\mu$m data and {\it{Herschel}} PACS data.

Because UV and MIPS derived SFRs are biased at large distances above the MS, as explained in Appendix B, we sample the SB region with PACS data. In addition, we use the large statistics of the COSMOS-PEP survey over 2 deg$^2$, to sample the rare SBs. Thus, in the intermediate and high redshift bins ($0.7 < z < 1.5$ and $1.5 < z <2.5$, respectively), where the GOODS$+$CANDELS sample is used, we combine this dataset with PACS detected subsample of the COSMOS-PEP dataset. The two datasets are volume-weighted before being compared. Fig. \ref{combo} shows the volume-corrected PACS selected COSMOS-PEP subsample with the orange histogram.

We check the shape of the residual distribution in the SB region by using the Bayesan Informative Criterion (BIC). The BIC is a rapid and robust method to check if a fitting function with a larger number of parameters is required to fit a dataset. The BIC can be expressed in the following form:
\begin{equation}
BIC={\chi}^2-k{\times}ln(n)
\end{equation}

where ${\chi}^2$ is the chi-squared, $k$ is the number of parameters and $n$ is the number of data points. The model with the lowest BIC is the best model. For this purpose, we fit the distributions of Fig. \ref{combo} per stellar mass bin with a single log-normal and a double log-normal function. In none of the redshift and stellar mass bins considered here, the BIC suggests that a second log-normal component is necessary to fit the upper envelope. Only in the lowest stellar mass bin at intermediate redshift ($0.7 < z < 1.5$), the BIC values are very close to each other, indicating that both fitting functions could be consistent with the data, although a single log-normal provides anyhow the lowest BIC value. In all other cases, the lowest BIC is obtained for a single log-normal distribution with high significance. The PACS selected COSMOS-PEP subsample in each bin is in agreement within 1$\sigma$ with being part of the SB tail of the single log-normal distribution. Such log-normal component is perfectly consistent within 1$\sigma$ with the rescaled local log-normal component. Thus, we conclude that the second log-normal component is not needed. 

As detailed in the Appendix B, the observation of a second log-normal component or an excess of galaxies in the SB region, appears to be due to the combination of different selection effects. The main bias is given by the combination of different SFR indicators. As shown in the Appendix B, both SFRs derived either from dust corrected UV luminosity or from {\it{Spitzer}} MIPS 24 $\mu$m are underestimated at large distances from the MS. This leads to an artificially tighter distribution in the upper envelope of the MS (right panel of Fig. \ref{app7} in Appendix B). Once PACS detections, which follow the actual distribution, are combined with the biased MIPS or UV based SFR distribution, they result as artificially overabundant.

In addition to this, the redshift dependence might lead to a similar effect, although at lower significance. If the redshift bin is very large or if there is strong evolution throughout it, neglecting the MS evolution within the bin leads to stretching the residual distribution in the upper envelope of the MS. Galaxies in the low and high redshift tail move artificially to larger values of \textit{$\Delta$MS}. This leads to a more leptokurtic distribution, very peaked at the center and with long tails, best fitted by two normal components (see Appendix B for further details). 

\cite{2015A&A...575A..74S} use a very similar dataset and find an excess of SB galaxies. However, this is observed when all \textit{$\Delta$MS} distributions, irrespective of the stellar mass bin, are combined together to increase the S/N. We point out that the increase of the MS scatter as a function of the stellar mass observed in \cite{2019MNRAS.483.3213P} and confirmed here up to $z\sim 2.2$, might cause this effect. Indeed, the combination of log-normal distributions with different dispersions leads to a leptokurtic distribution where the component with the smaller dispersion and higher amplitude dominates the peak, while the component with larger dispersion and lower amplitude dominates the tails.

\begin{figure*}
\includegraphics[width=\columnwidth]{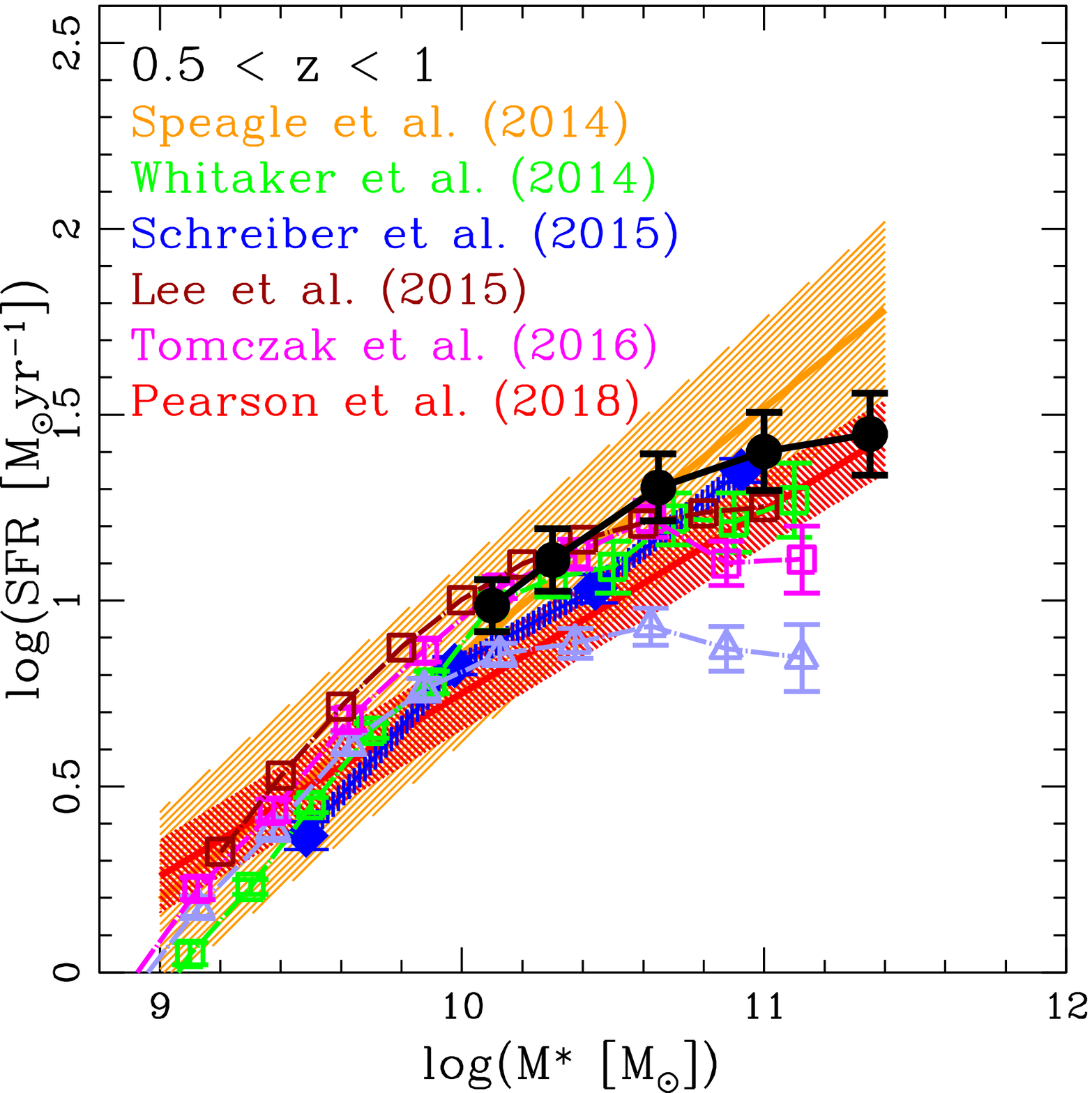}
\includegraphics[width=\columnwidth]{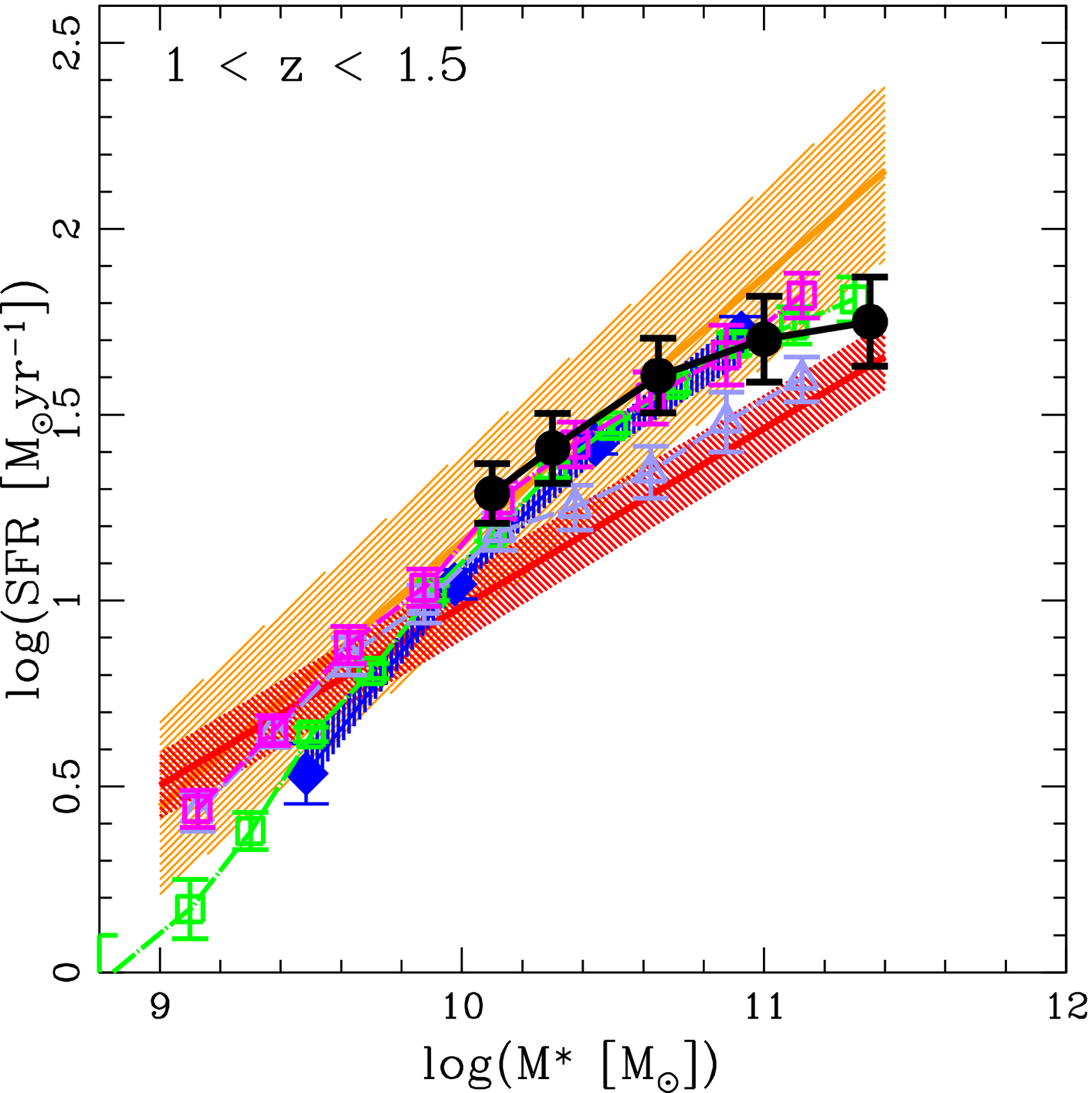}
\includegraphics[width=\columnwidth]{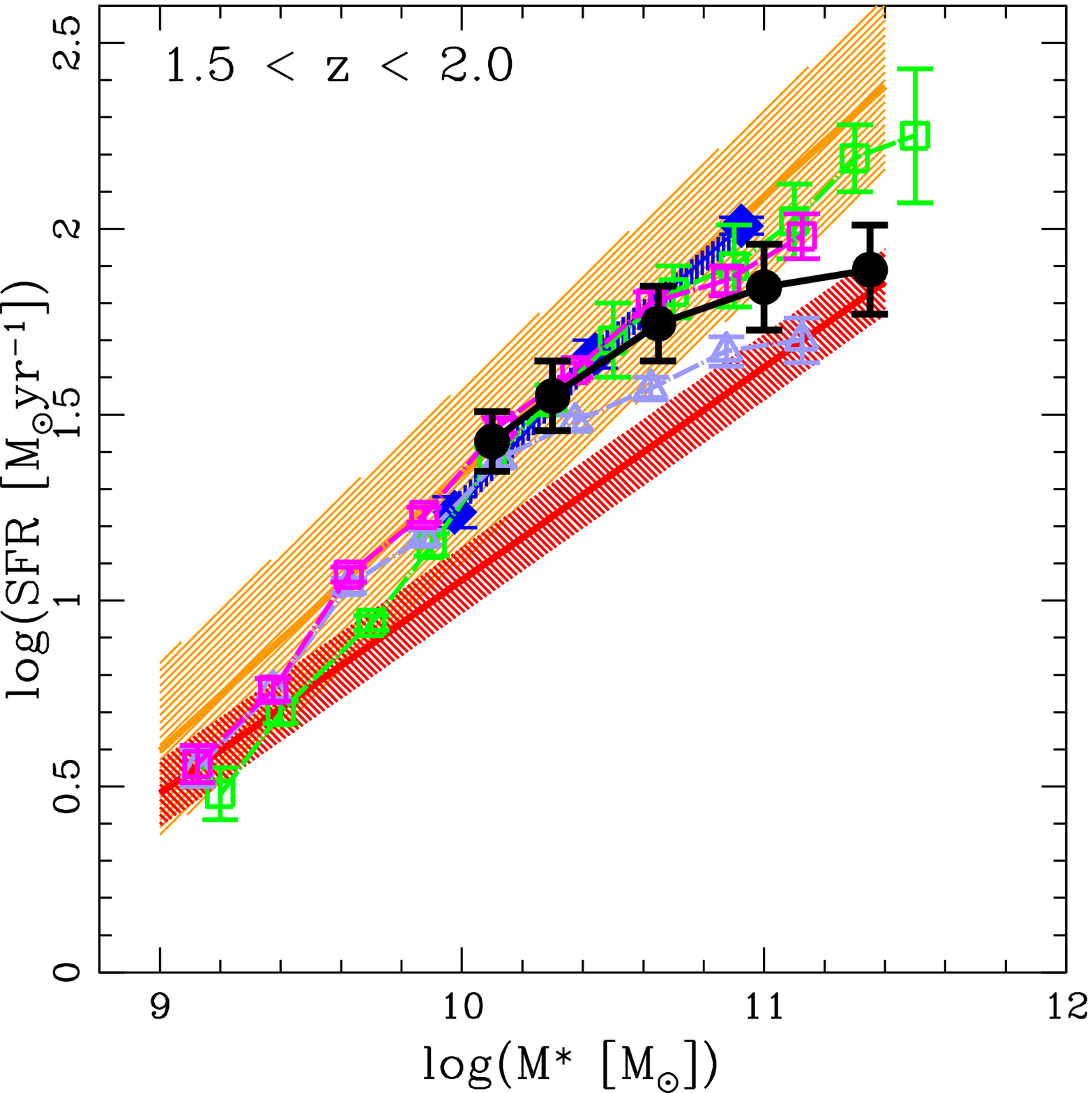}
\includegraphics[width=\columnwidth]{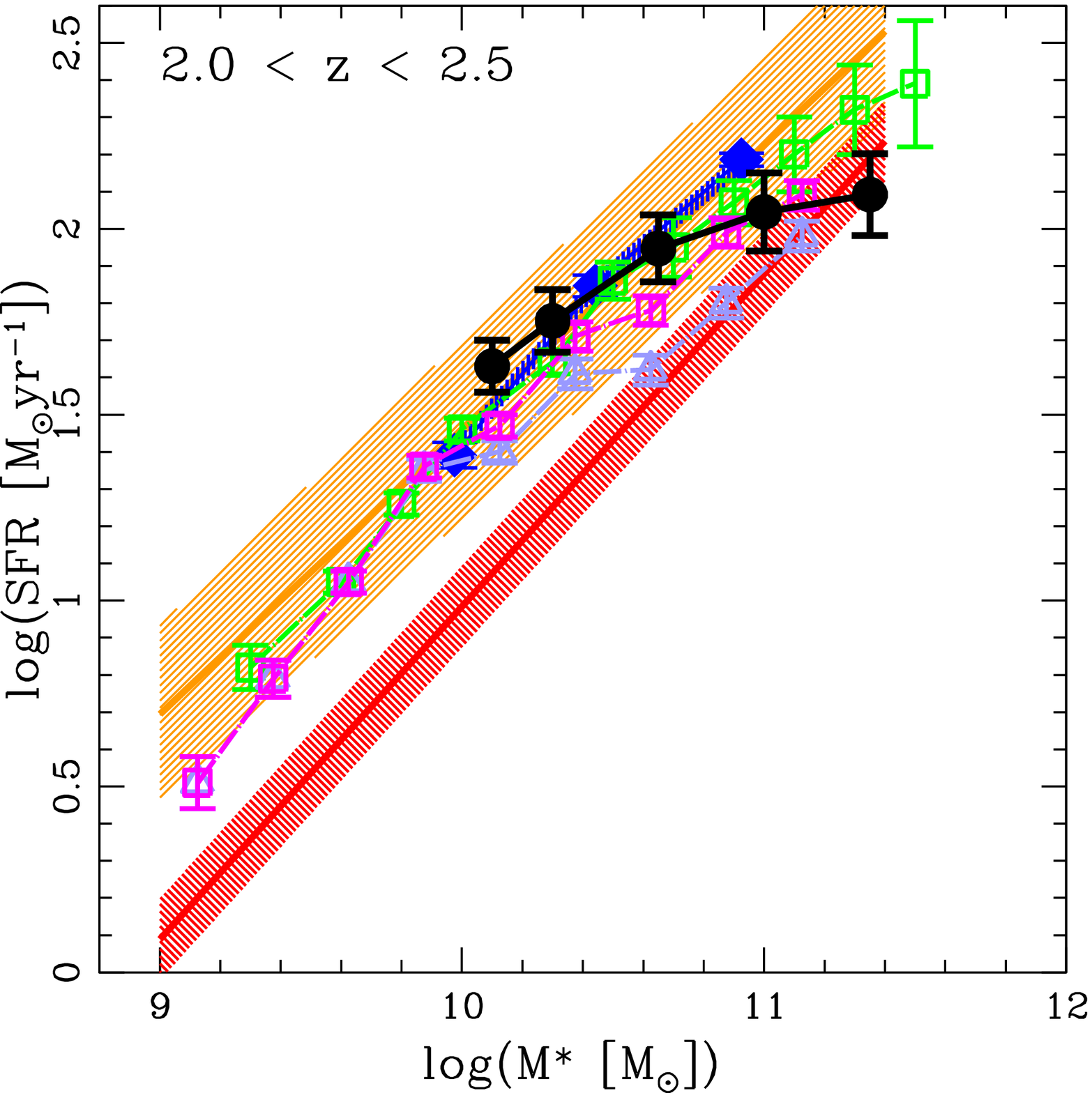}
\caption{Comparison of the MS location obtained in this work (filled points) with previous results, in four redshift bins ($0.5<z<1$, $1<z<1.5$, $1.5<z<2$ and $2<z<2.5$). The redshift windows are chosen to match previous works. The symbols and lines are color-coded as indicated in the figure. The shaded regions and the error bars of each MS are the 1$\sigma$ uncertainties reported in the literature. The light violet triangles in the panels show the results of the stacking analysis of \protect\cite{2016ApJ...817..118T} based on all galaxies without any SFG pre-selection.}
\label{comparison}
\end{figure*}

\begin{figure*}
\includegraphics[width=17.cm, height=13.cm]{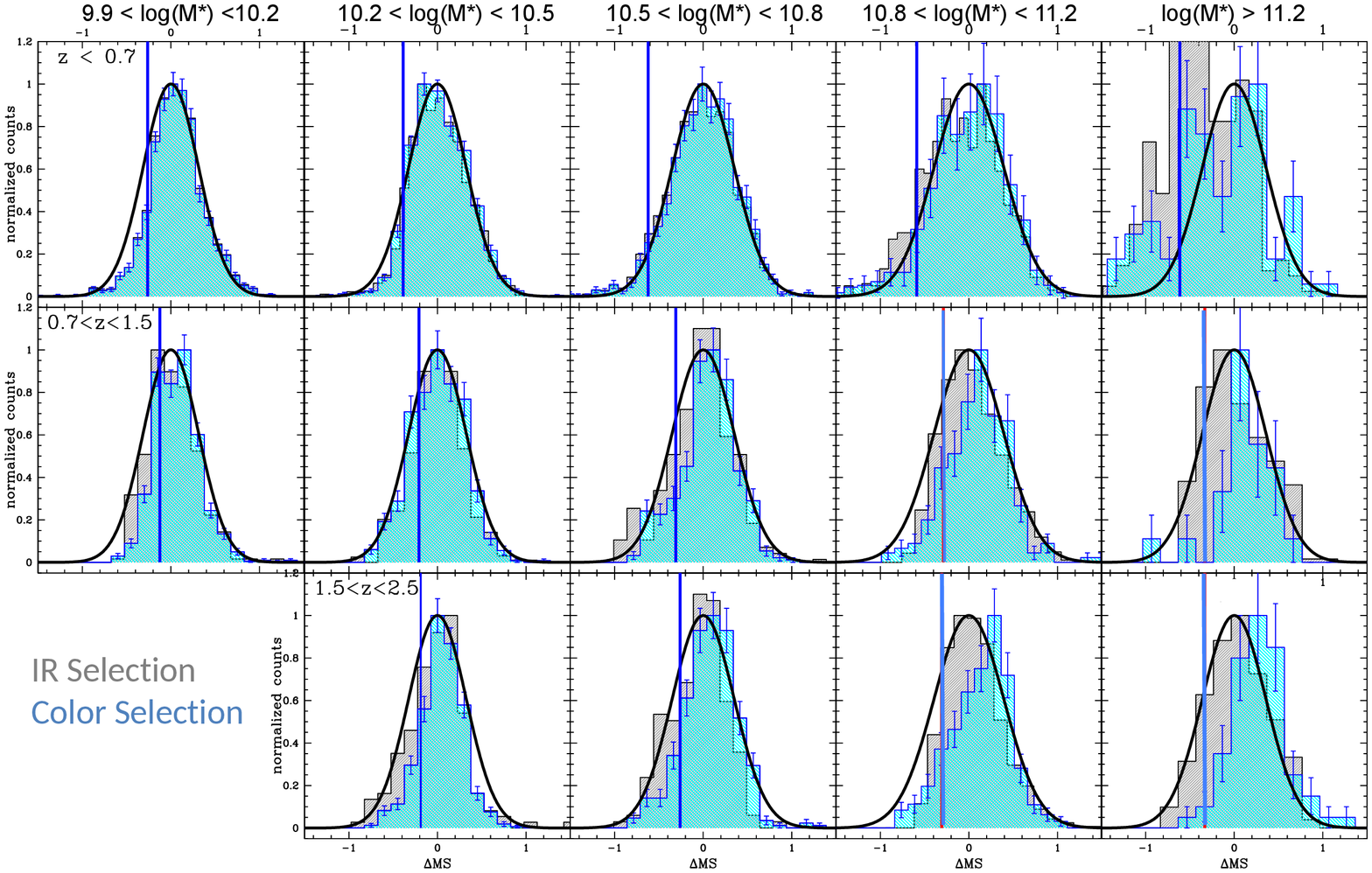}
\caption{Comparison of the distribution of galaxies around the MS for purely IR selected galaxies (grey histograms) and IR and color-selected galaxies (cyan histograms), according to the color selection criteria of \protect\cite{2015A&A...575A..74S}. The redshift and stellar mass bins are the same as in Fig. \ref{combo}.  The blue vertical lines show the 80\% completeness limit of the IR selected sample as in Fig \ref{combo}. }
\label{ecco}
\end{figure*}

\section{Comparison with previous works}
Here we compare our results with robust determinations of the MS at the high mass end, available in literature. We also describe in detail the reasons of agreement and disagreement with each determination.

We consider the best fit provided by \cite{2014ApJS..214...15S} as a good statistical representation of all the results available in literature before 2014 \cite[25 MS determinations as listed in][]{2014ApJS..214...15S}. We consider, in particular, the best fit n. 49, which is indicated as the reference best fit by \cite{2014ApJS..214...15S}. This is based on a "mixed" selection of SFGs, mainly driven by color-color selection of different kinds and cuts \cite[see discussion in][]{2014ApJS..214...15S}. We consider also the results of \cite{2014ApJ...795..104W}, \cite{2015A&A...575A..74S} and \cite{2016ApJ...817..118T}, who perform the stacking of SFGs in {\it{Spitzer}} MIPS and {\it{Herschel}} PACS maps. In these works, SFGs are pre-selected in a similar way on the basis of the U-V$-$V-J rest-frame colors. We consider the MS of \cite{2015ApJ...801...80L}, based on a ladder of SFR indicators (dust corrected UV, MIPS 24 $\mu$m and {\it{Herschel}} PACS) and fitted with power law with a mass turn-over. In this work SFGs are selected according to their ($M_{NUV}$-$M_R)-(M_R$-$M_J$) absolute color as in \citet[][2015]{2013A&A...556A..55I}. We include in the comparison the more recent results of Pearson et al. (2018) based on the deblending of {\it{Herschel}} SPIRE detections. In this case, SFGs are selected with similar color cuts. We do not consider the work of \cite{2016ApJ...820L...1K} and \cite{2017ApJ...847...76S}, based on dust corrected UV luminosities, because they include only few tens of galaxies in the mass range $10^{10}-10^{11}$ $M_{\odot}$. Therefore, they can not provide a robust determination of the MS slope at the high mass end. All MS are converted to a Chiabrier IMF. 

The comparison is shown in Fig. \ref{comparison}. The MS of \citet[orange shaded region]{2014ApJS..214...15S} and \citet[][red shaded region]{2018A&A...615A.146P} do not show any bending of the MS at any redshift. However, the MS determination of \cite{2014ApJS..214...15S} exhibits a very large uncertainty with respect to other determinations. This likely reflects the large spread in the results of the 25 MS determinations collected in that analysis. For this reason the MS of \cite{2014ApJS..214...15S} is consistent within 1$\sigma$ with all other determinations, except \cite{2018A&A...615A.146P}. As already pointed out previously, \cite{2018A&A...615A.146P} report a systematic offset of their SFR of $\sim$0.4 dex below previous results. This explains why their MS lies below the other determinations at more than 1$\sigma$ in 3 out of 4 redshift bins. \cite{2011A&A...533A.119E} show that SPIRE and PACS SFR estimators lead to consistent results. So the discrepancy must be related to the deblending technique of the SPIRE detections and the SED fitting technique applied in \citet[][see their Appendix C for an extensive discussion]{2018A&A...615A.146P}.

The MS determinations of \citet[][green squares]{2014ApJ...795..104W}, \citet[][blue shaded region]{2015A&A...575A..74S} and \citet[][magenta squares]{2016ApJ...817..118T} show a bending of the MS below $z \sim 1.5$, and a steeper relation at higher redshift. We point out that the stacking analysis provides, by construction, the mean IR flux of the selected sample. This is used to estimate the mean SFR of SFG population. A median flux can also be estimated, but it is not consistent with the median of the SFR distribution. Thus, to compare our results with those of the stacking analysis, we use the MS location based on the mean SFR shown in Fig. \ref{riassunto} (blue points), shifted to high redshift according to Eq. 2. 

As shown in Fig. \ref{comparison}, our results (black points) and the MS based on the stacking are in perfect agreement (within 1$\sigma$) up to $z \sim 1.5$. Above this threshold we observe less agreement (within $1.5\sigma$) in particular above $10^{11}$ $M_{\odot}$. To test what might cause this small discrepancy, we investigate the effect of the color selection used to identify SFGs, with respect to our IR selection. Fig. \ref{ecco} shows in gray the $\Delta{MS}$ distribution of the COSMOS-PEP and GOODS$+$CANDELS sample in the same redshift and stellar mass bins of Fig. \ref{combo}. The cyan histograms show, instead, the distribution of galaxies selected according to the (U-V$)-($V-J) color criteria of \cite{2015A&A...575A..74S} applied to the IR selected sample. Fig. \ref{ecco} shows clearly that the color selection is able to capture all the IR selected MS galaxies up to $10^{10.5-10.8}$ $M_{\odot}$ and it does not affect the shape of the galaxy distribution in the MS region. Thus, the mean SFR resulting from the stacking analysis is consistent with the mean SFR derived from the log-normal distribution, as shown in Fig. \ref{comparison}. At higher stellar masses, we observe a redshift dependent effect. At low redshift ($0 < z < 0.7$), the color selection leads to a $\Delta${MS} distribution consistent with our best fit log-normal distribution. At higher redshift ($z > 0.7$) the same selection depopulates the lower envelope of the distribution, thus leading to a larger mean SFR, as observed in Fig. \ref{comparison}. This is because, at $z > 0.7$ the color selection captures 70\% of the IR selected galaxies above $10^{10.5}$ $M_{\odot}$ and only 50\% above $10^{10.8}$ $M_{\odot}$, those with the bluer color on the upper envelope of the MS. This is observed also in \cite{2015A&A...575A..74S}. However, $10^{11}$ $M_{\odot}$ galaxies at 1-2$\sigma$ below the peak of the log-normal distribution exhibit a specific SFR of $10^{-9.3}-10^{-9.6}$ $yr^{-1}$ at $z\sim 2$ and $10^{-9.8}-10^{-10.1}$ $yr^{-1}$ at $z\sim 1$, respectively. This high level of SF activity is not consistent with a quiescent system, neither is the high FIR flux density consistent with emission of low mass stars. 
The same result is observed also when using the color selection criteria and the MIPS based SFRs of \cite{2014ApJ...795..104W}. \cite{2016ApJ...817..118T} explores the effect of the SFG pre-selection by stacking all galaxies irrespective of their colors. The relation obtained in this way (light violet triangles in Fig. \ref{comparison}) lies well below all other determinations, including those obtained here, at all redshifts and masses. This shows that our method is able to isolate the MS location with respect to other loci without introducing additional selection biases. We conclude that the previously reported steepening of the MS slope towards higher redshift is artificially induced by the color selection. Consequently, so is also the evolution of the mass turn-over as observed in \cite{2015ApJ...801...80L}, \cite{2015A&A...581A..54T} and \cite{2016ApJ...817..118T}. This selection effect likely affects also the MS of \cite{2014ApJS..214...15S}, which is based on a "mixed" selections. However, differently from the stacking analysis, the statistical study of \cite{2014ApJS..214...15S}, which combines and calibrates different SFR indicators and samples, does not allow to estimate to which extent such bias affects the result.

We conclude that the MS is not evolving in slope at the high mass end and the relation is bending above $10^{10.5}$ $M_{\odot}$, as observed in the local Universe. Since there is agreement in literature in finding that the MS is marginally evolving in slope at the low mass end, this would suggest that the shape of the relation is not evolving up to z$\sim$2.5 over a much broader stellar mass range.

\section{Summary and Conclusion}

We summarize here the main findings of the paper. We study the MS of SFGs from z$\sim$0 to $\sim$2.5 by analyzing the SFR distribution in the SFR-M$_*$ plane. To this aim, we use the deepest available mid- and FIR surveys of the major blank fields, such as COSMOS, GOODS and CANDELS. 

To study the evolution of the MS location and shape at high redshift, we assume as a null hypothesis that the MS slope and scatter do not evolve with time and only the normalization of the relation increases with redshift.  We find that the null hypothesis is confirmed out to redshift $\sim$ 2.5 with the exclusion of the highest stellar mass bin ($>10^{11.2} M_{\odot}$), where the results of the test are inconclusive. Given the validity of the null hypothesis, we conclude that the MS is bending above $10^{10.5}$ $M_{\odot}$ out to z$\sim$2.5, consistently with the local MS.  We show that previously reported steepening of the MS towards high stellar masses is due to a selection effect. 

The distribution of galaxies in the MS region, at fixed stellar mass, is well represented by the local log-normal distribution observed in \cite{2019MNRAS.483.3213P}. We conclude that, up to $z\sim2.5$, the distribution of galaxies in the MS region is consistent with being a re-scaled version of the local distribution, whose scatter increases as a function of the stellar mass at the $2\sigma$ level. We also show that SBs represent the high SFR tail of the log-normal distribution and they are not in excess with respect to the underlying SFG distribution, as previously proposed in the literature \citep[e.g.][]{2012ApJ...747L..31S}.

\section*{Acknowledgements}
This research was supported by the DFG cluster of excellence "Origin and Structure of the Universe" (www.universe-cluster.de). P.P. thanks A. Renzini, M. Sargent and E. Daddi for the very useful discussions.




\bibliographystyle{mnras}
\bibliography{MS_hz_new} 



\appendix

\section{Comparison of different SFR indicators}

\begin{figure}
\includegraphics[width=\columnwidth]{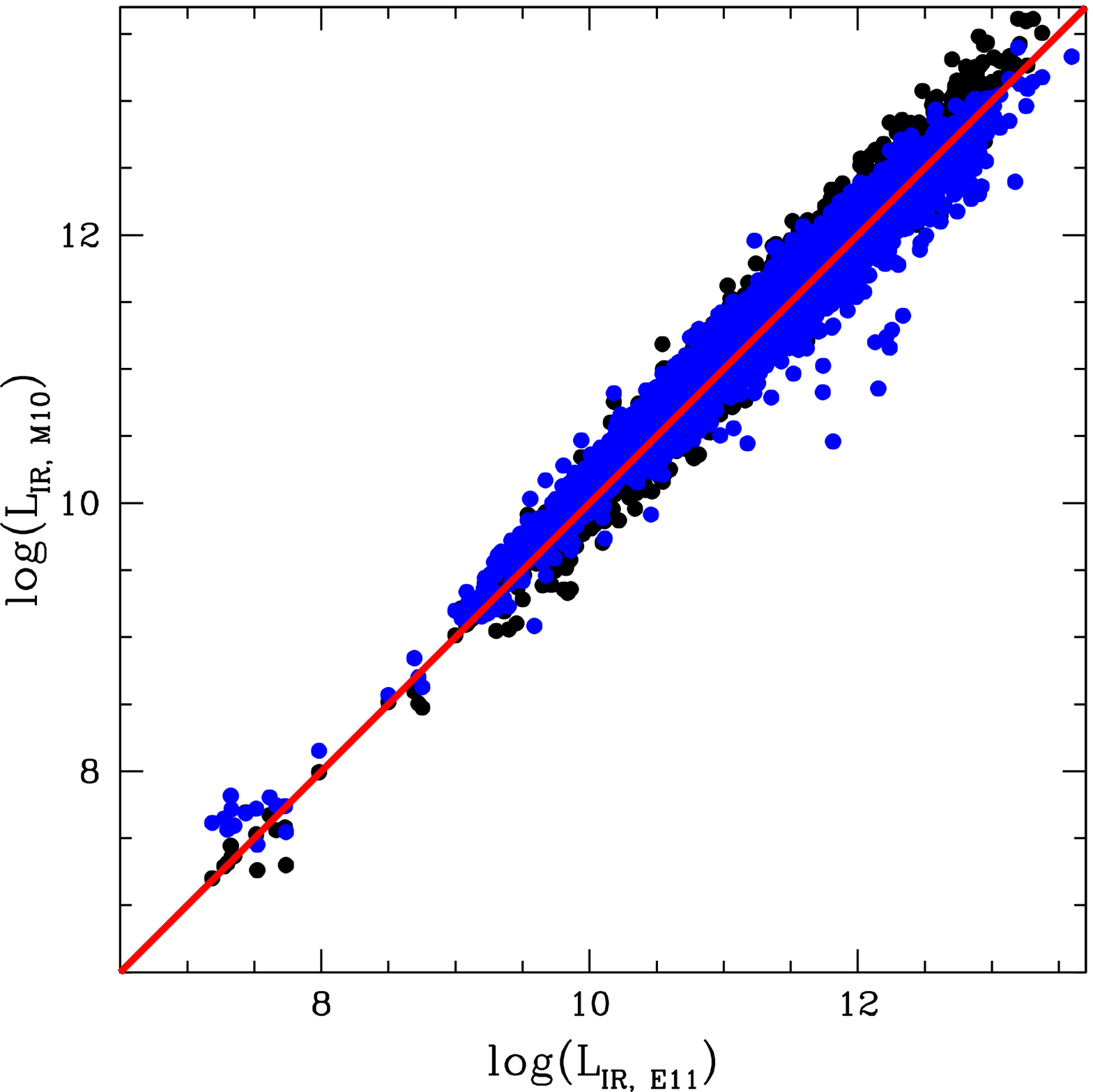}
\caption{ The panel shows the comparison of the $L_{IR}$ derived by using the \protect\cite{2011A&A...533A.119E} IR templates versus those based on the \protect\cite{2010ApJ...714.1740M} templates (black points). The panel shows also the comparison between the \protect\cite{2011A&A...533A.119E} $L_{IR}$ and those derived from fitting the whole galaxy SED with {\it{hyperz}} with \protect\cite{2003MNRAS.344.1000B} templates in the optical regime and the \protect\cite{2007ApJ...663...81P} templates in the IR regime (blue points). The $L_{IR}$ is obtained by integrating the template best fitting the MIPS and PACS data from 8 to 1000 $\mu$m}
\label{app2}
\end{figure}

In this section we compare SFRs derived with different indicators to highlight possible biases. First, we compare the IR luminosities, $L_{IR}$, derived by integrating different templates from 8 to 1000$\mu$m. Fig. \ref{app2} shows the comparison of $L_{IR}$ obtained by using the \cite{2011A&A...533A.119E} and the \cite{2010ApJ...714.1740M} templates, respectively. The agreement is excellent with an rms of 0.16 dex. The same panel shows also the comparison of the $L_{IR}$ obtained by fitting simultaneously with {\it{hyperz}} the best optical \citep{2003MNRAS.344.1000B} and IR \citep{2007ApJ...663...81P} templates to the entire galaxy SED from UV to PACS data (blue points). The agreement is very good with a rms of 0.17 dex.

\begin{figure*}
\includegraphics[width=\columnwidth]{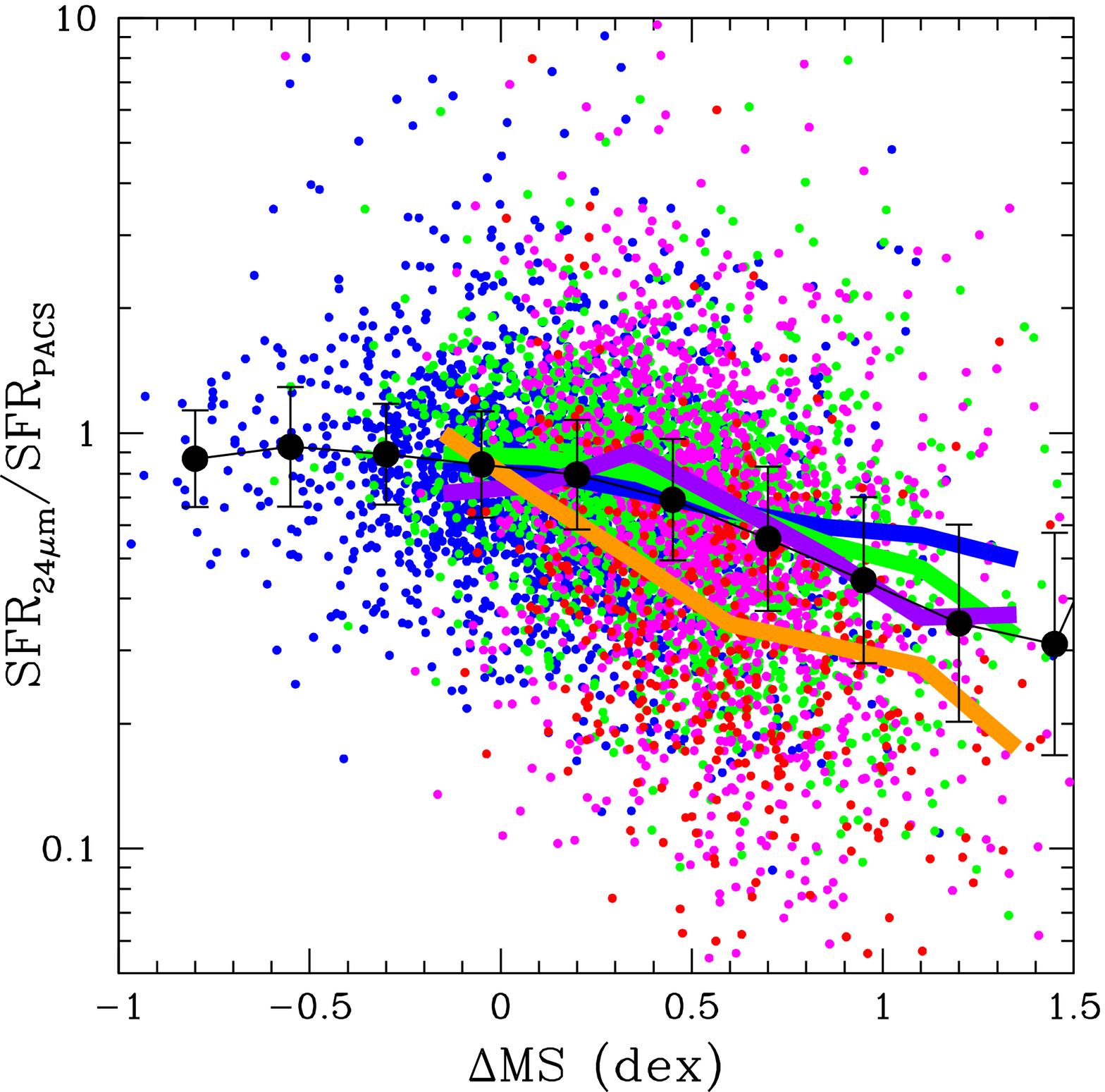}
\includegraphics[width=\columnwidth]{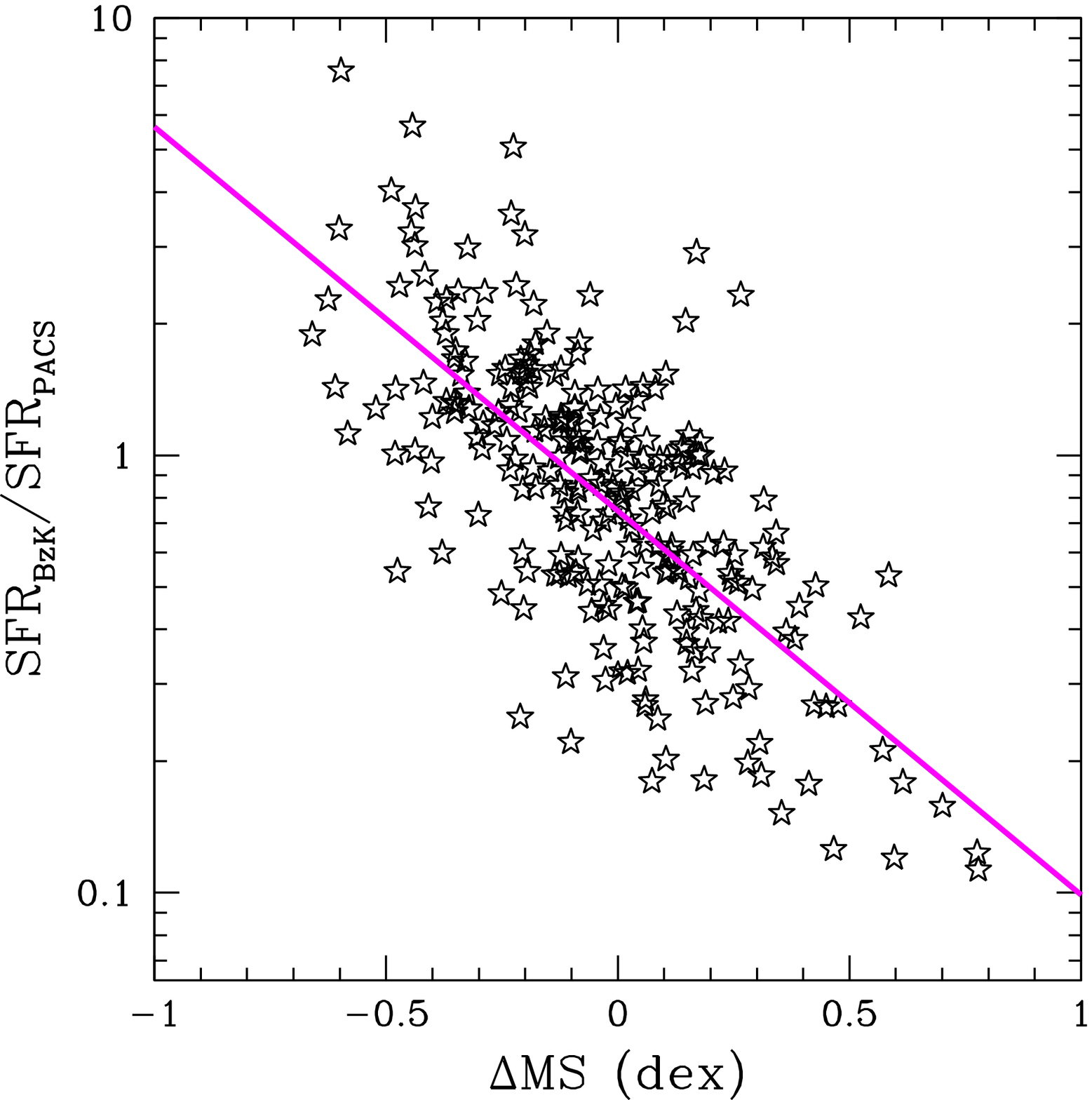}
\caption{{\it{Left panel}}: relation between the ratio, $SFR_{24{\mu}m}/SFR_{PACS}$, of SFRs based on 24 $\mu$m data-point only and on 24 $\mu$m$+$PACS data, versus the distance from the MS of \protect\cite{2014ApJ...795..104W} at different redshifts. Blue, green, magenta and red points refer to the $0 < z < 0.5$, $0.5 < z < 1$, $1 < z < 1.8$ and $1.8 < z < 2.5$  redshift ranges, respectively, in the COSMOS PEP sample. The black dots shows the mean $SFR_{24{\mu}m}/SFR_{PACS}$ in bin of $\Delta{MS}$ for the whole COSMOS PEP sample. The blue, green, purple and orange line show the mean relation at $0 < z < 0.5$, $0.5 < z < 1$, $1 < z < 1.8$ and $1.8 < z < 2.5$, respectively. {\it{Right panel}}:  ratio, $SFR_{BzK}/SFR_{PACS}$, of the dust corrected UV based SFRs of the BzK sample of \protect\cite{2011ApJ...739L..40R} at $1.5 < z < 2.5$ over the PACS based SFRs, versus the distance from the MS. The comparison is done with the PACS based SFRs derived from the deep CANDELS and GOODS-H {\it{Herschel}} PACS data. The magenta line shows the best fit linear regression of the $SFR_{BzK}/SFR_{PACS}$ relation.}
\label{app7}
\end{figure*}

However, when comparing the SFR based on MIPS 24 $\mu$m data-point only, with the PACS based SFR, all templates exhibit the same bias towards the upper envelope of the MS. At all redshifts, all the templates considered here tend to underestimate the SFR based on 24 $\mu$m towards the upper envelope of the MS. The larger the distance from the MS, the larger the underestimation. We also note that the higher the redshift, the steeper the anti-correlation between the $SFR_{MIPS}/SFR_{PACS}$ and $\Delta{MS}$ (left panel of Fig. \ref{app7}). This is due to the fact that at higher redshift the MIPS 24 $\mu$m data sample a rest-frame region more contaminated by PAHs (the rest frame 12 $\mu$m region at $z\sim 1$ and the 8 $\mu$m region at $z \sim 2$) than the rest frame 24 $\mu$ region. The PAHs emission appears to be lower for galaxies above the MS than for MS galaxies, as shown in\cite{2011A&A...533A.119E}. Thus, when using the MS template also for SBs, the $L_{IR}$ estimated from the continuum emission is underestimated with respect to the actual value. Only PACS data at longer wavelength allow to discriminate between a MS or a SB template to properly estimate the infrared luminosity.

We perform the same exercise with the dust corrected UV based SFRs of the BzK sample of \cite{2011ApJ...739L..40R} at $1.5 < z < 2.5$ (right panel of Fig. \ref{app7}). The UV based SFRs of the BzK sample behave in a very similar way with respect to the local galaxy sample of Salim et al. (2016), analysed in the very same way in \cite{2019MNRAS.483.3213P}. We observe a significant underestimation of the SFR at large distances from the MS. As for the left panel of Fig. \ref{app7}, we use the MS determination of \cite{2014ApJ...795..104W} at different redshifts to estimate the distance from the MS.

\begin{figure*}
\includegraphics[width=\columnwidth]{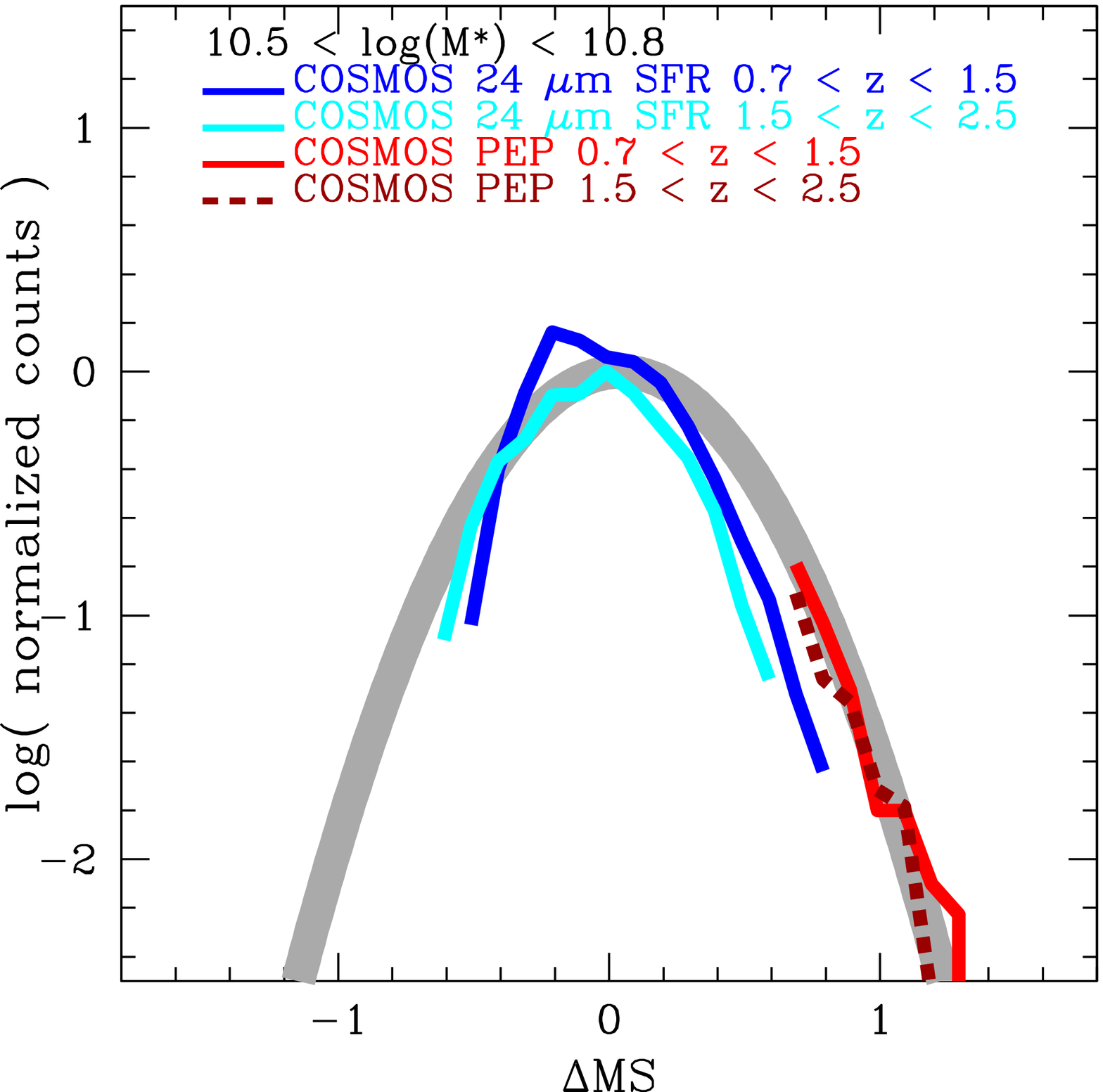}
\includegraphics[width=\columnwidth]{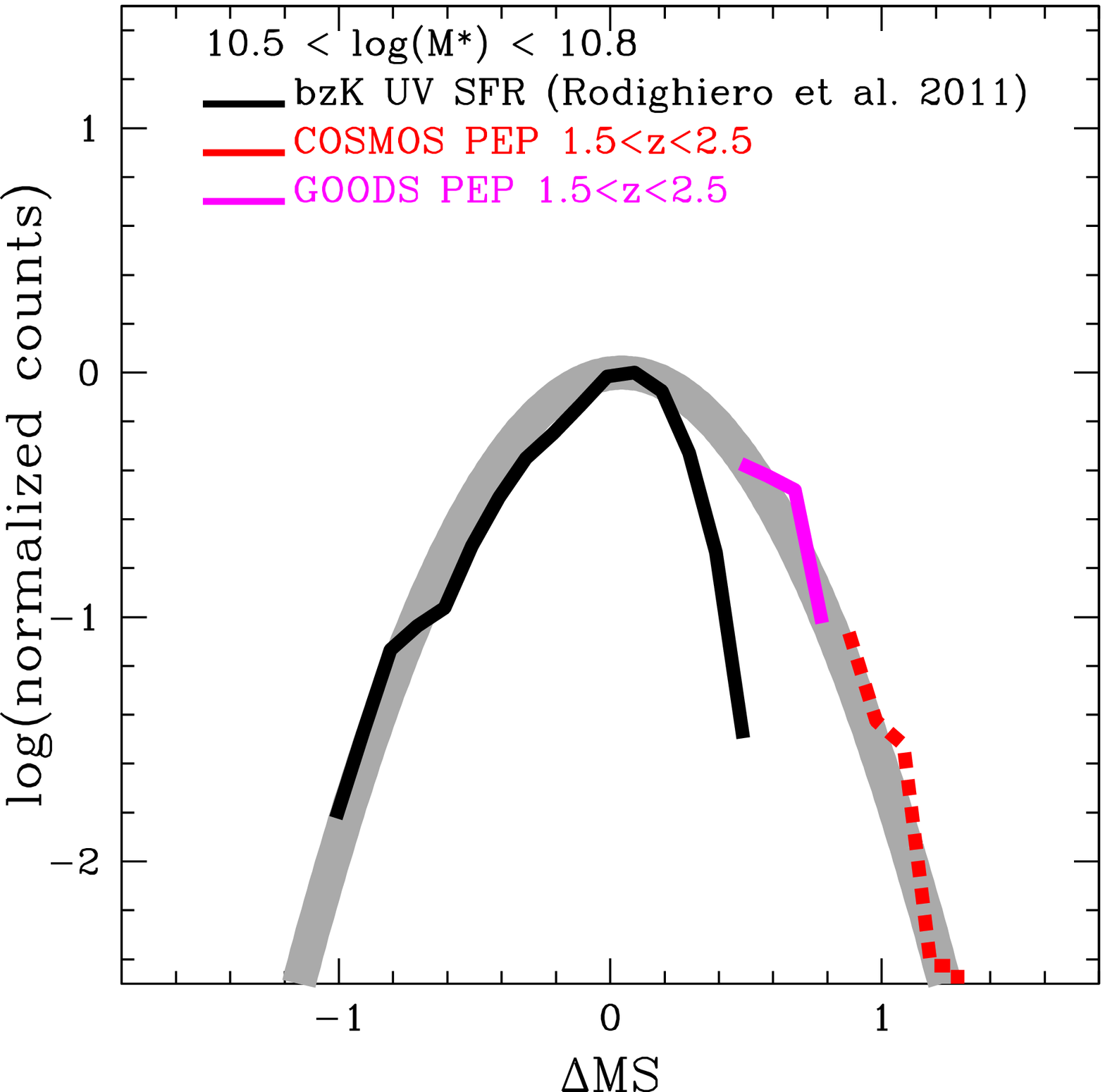}
\caption{{\it{Left panel}}: Bias due to the combination of different SFR indicators in the SFR distribution around the MS at $10^{10.5-10.8}$ $M_{\odot}$. Cyan and blue histograms shows the distribution of MIPS 24 $\mu$m sources in the COSMOS-PEP field at different redshift, as reported in the figure. The red and dark red histograms, in particular, shows the subsample of PACS detected sources of the COSMOS-PEP sample. The gray tick line shows the $z\sim0$ best fit relation of \protect\cite{2019MNRAS.483.3213P} in the same stellar mass bin. {\it{Right panel}}: Same as left panel for a different combination of SFR indicators. The black curve shows log(SFR) distribution of the bzK sample of \protect\cite{2011ApJ...739L..40R} with SFRs based on the dust corrected UV luminosities (as shown in right panel of Fig. \ref{app7}). The magenta and red lines show the distributions of the PACS selected samples in the COSMOS PEP and GOODS PEP surveys, respectively. The gray tick line shows the $z\sim0$ best fit relation of \protect\cite{2019MNRAS.483.3213P} in the same stellar mass bin}
\label{bias}
\end{figure*}

\begin{figure}
\includegraphics[width=\columnwidth]{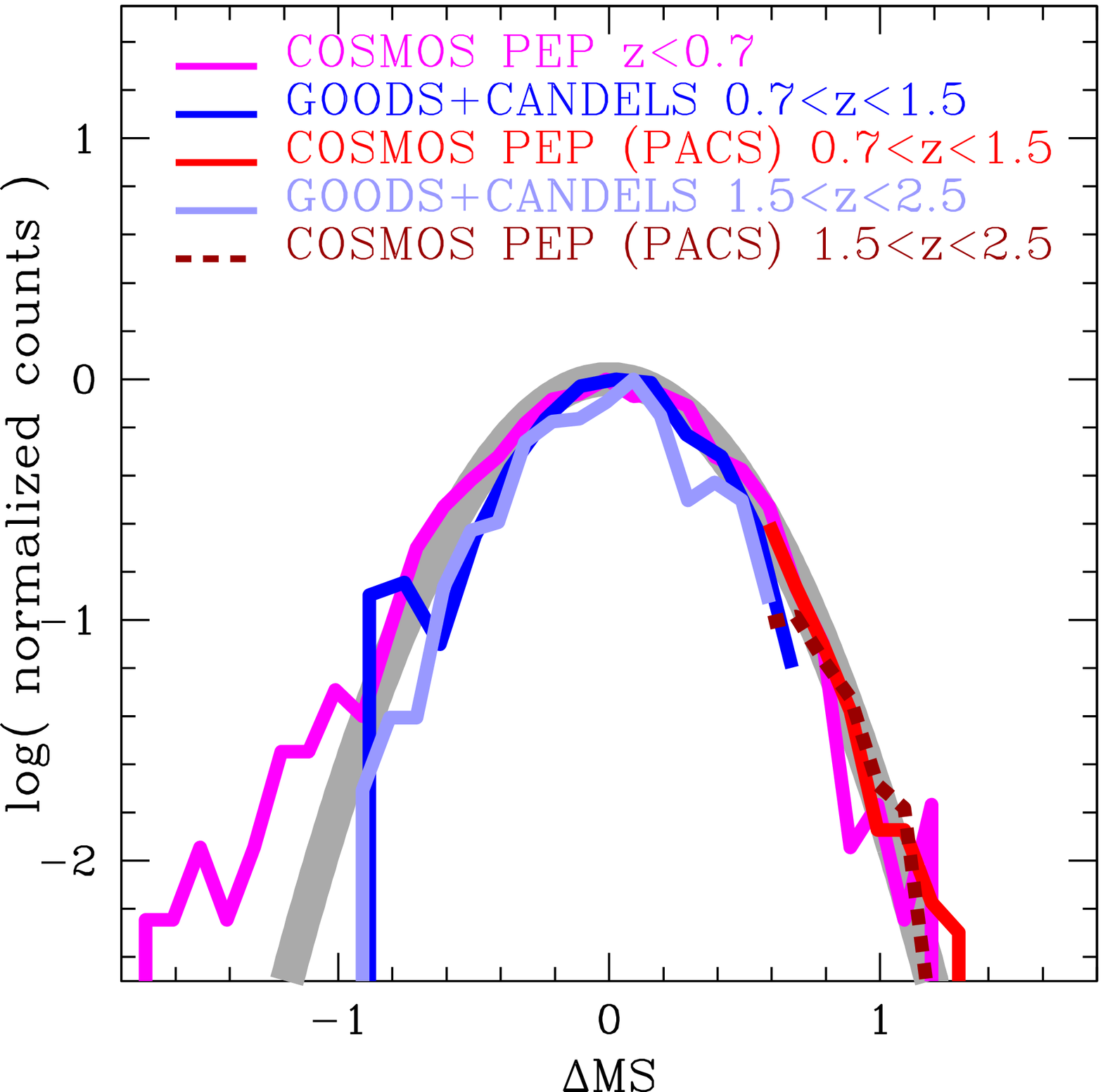}
\caption{Example of the distribution of galaxies around the MS location at different redshift in the $10^{10.5-10.8}M_{\odot}$ stellar mass, once the upper envelope is fully sampled by PACS data as in the GOODS$+$CANDELS sample. The magenta line shows the distribution of the COSMOS-PEP sample at $z < 0.7$. The blue and cyan lines show the GOODS$+$CANDELS sample at $0.7 < z < 1.5$ and $1.5 < z < 2.5$, respectively. The red solid and dashed histogram, in particular, show the volume weighted subsample of PACS detected sources of the COSMOS-PEP sample at $0.7 < z < 1.5$ and $1.5 < z < 2.5$. All histograms are in agreement with the best fit $z \sim 0$ log-normal distribution of \protect\cite[][gray thick line]{2019MNRAS.483.3213P} within 1$\sigma$.}
\label{SB}
\end{figure}

\begin{figure}
\includegraphics[width=\columnwidth]{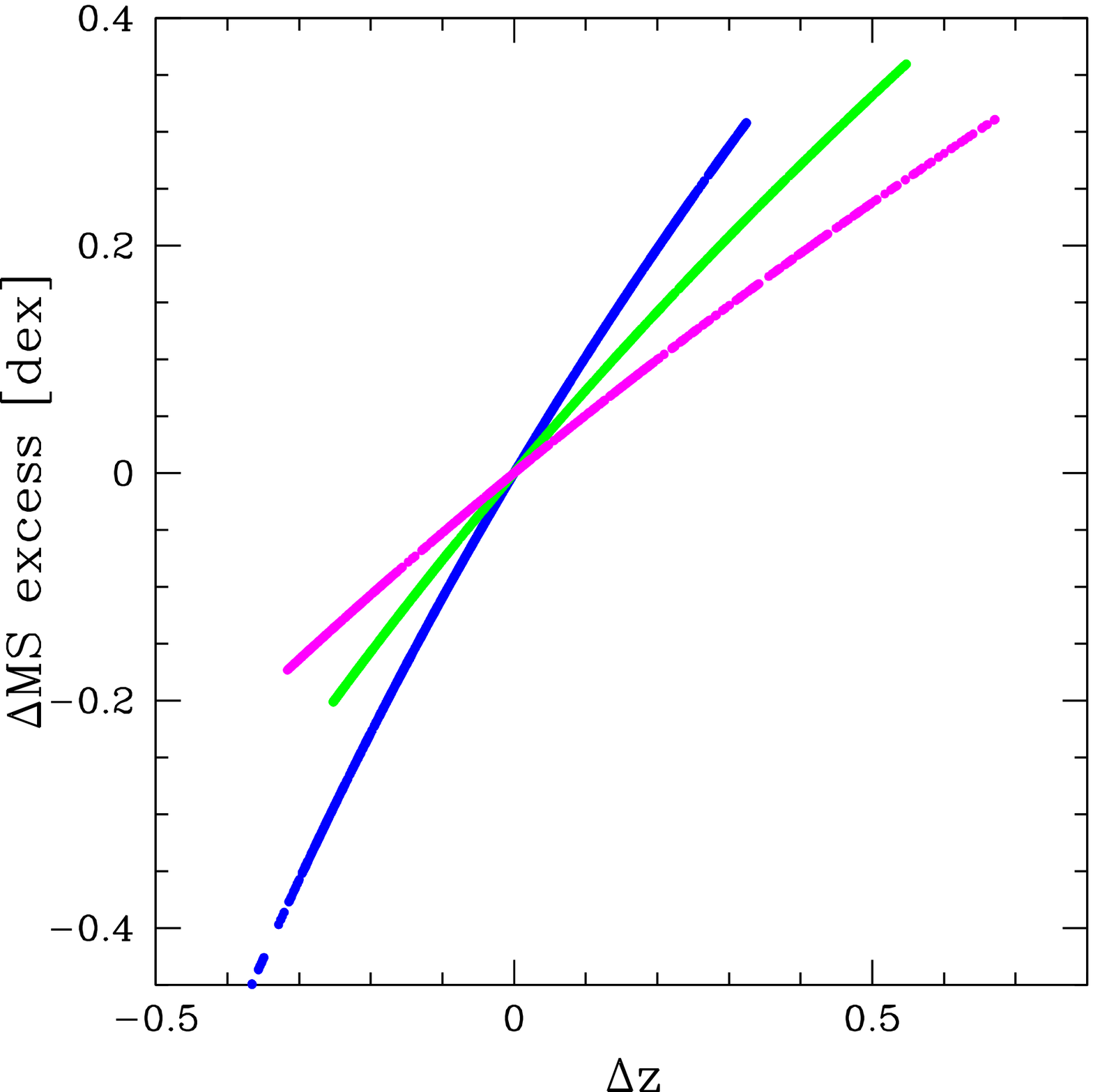}
\caption{Excess of $\Delta{MS}$ in three bins of redshift at $0 < z < 0.7$ (blue region), $0.7 < z < 1.5$ (green region) and $1.5 < z < 2.5$ (blue region) due to neglecting the evolution of the MS location within wide redshift bins.}
\label{ev_bias}
\end{figure}

\section{Biases in the MS shape determination}
In this section we analyze the possible biases that can affect the determination of the SFR distribution around the MS. Given the results of the previous section, the most obvious bias is given by the combination of different SFR indicators. The use of the SFR based on the on 24 $\mu$m data-point only tend to depopulate the high SFR tail of the MS due to the underestimation of the SFR in the SB region. In the SB region, in particular, such underestimation can be of 0.3 to 0.7 dex from redshift $\sim$ 1 to 2.5, respectively (Left panel of Fig. \ref{app7}), leading to a very narrow MS at all stellar masses when using only MIPS 24 $\mu$m SFR indicator. The same bias is observed in the SFR estimates based on the dust corrected UV luminosities of the BzK sample of \cite{2011ApJ...739L..40R}. 

This systematic underestimation of the SFRs at larger distance from the MS could have few effects. First, it would artificially create a narrow MS at all masses, likely hiding the increase of the scatter of the relation towards high stellar masses, as observed in Fig. \ref{riassunto} and \ref{combo}. Second, the combination of such biased SFR estimates with the PACS based SFRs in the upper envelope would lead to an artificial leptokurtic distribution. As shown in Fig. \ref{bias}, the PACS detected galaxies in the SB region appear as artificially overabundant with respect to the 24 $\mu$m based (left panel) and dust corrected UV based (right panel) SFR distributions. However, once a larger part of the MS upper envelope is sampled by PACS data, as in the GOODS$+$CANDELS sample, such artificial effect disappears (Fig. \ref{SB}).

Additionally, the MS location is evolving strongly as a function of redshift at the high mass end with a power law of the form $(1+z)^{3.21}$, as found in Section 3.2. Ignoring such evolution, within a large redshift bin, affects the shape of the MS as shown in Fig. \ref{ev_bias}. Galaxies towards the lower and upper limit of the redshift bin, are located to artificially smaller and larger distances, respectively, from the MS of the bin mean redshift. The result is that the MS is squeezed in the lower envelope and stretched in the upper envelope with an artificially long tail.


\bsp	
\label{lastpage}
\end{document}